\documentclass[preprint,12pt]{elsarticle}

\usepackage{graphicx}
\usepackage{amsmath}
\usepackage{commath}
\usepackage{mathtools}
\usepackage{amssymb}
\usepackage{lineno,hyperref}
\usepackage{algorithm}
\usepackage{algpseudocode}

\usepackage{subfigure}
\usepackage{bm}
\usepackage{cleveref}

\usepackage{graphicx}
\usepackage{booktabs}
\usepackage{siunitx}
\begin{document}

\begin{frontmatter}

%% Title, authors and addresses

%% use the tnoteref command within \title for footnotes;
%% use the tnotetext command for theassociated footnote;
%% use the fnref command within \author or \address for footnotes;
%% use the fntext command for theassociated footnote;
%% use the corref command within \author for corresponding author footnotes;
%% use the cortext command for theassociated footnote;
%% use the ead command for the email address,
%% and the form \ead[url] for the home page:
%% \title{Title\tnoteref{label1}}
%% \tnotetext[label1]{}
%% \author{Name\corref{cor1}\fnref{label2}}
%% \ead{email address}
%% \ead[url]{home page}
%% \fntext[label2]{}
%% \cortext[cor1]{}
%% \affiliation{organization={},
%%             addressline={},
%%             city={},
%%             postcode={},
%%             state={},
%%             country={}}
%% \fntext[label3]{}

\title{Metaball-Imaging Discrete Element Lattice Boltzmann Method for fluid-particle system of complex morphologies with settling case study}

%% use optional labels to link authors explicitly to addresses:
%% \author[label1,label2]{}
%% \affiliation[label1]{organization={},
%%             addressline={},
%%             city={},
%%             postcode={},
%%             state={},
%%             country={}}
%%
%% \affiliation[label2]{organization={},
%%             addressline={},
%%             city={},
%%             postcode={},
%%             state={},
%%             country={}}

\author[inst1,inst2,inst3]{Yifeng Zhao}

\author[inst2,inst3]{Pei Zhang}
% \author[inst3]{Xiangbo Gao}

\author[inst3]{Liang Lei}

\author[inst2,inst3]{S.A. Galindo-Torres\corref{cor2}}
\tnotetext[label1]{\\}
\cortext[cor1]{Corresponding author,  Tel: +86 15158184891}
\ead{s.torres@westlake.edu.cn}

\author[inst3]{Stan Z. Li\corref{cor}}
% \cortext[cor]{Stan.ZQ.Li@westlake.edu.cn}

\affiliation[inst1]{organization={Collage of Environmental and Resources Science, Zhejiang University},%Department and Organization
            addressline={866 Yuhangtang Road}, 
            city={Hangzhou},
            postcode={310058}, 
            state={Zhejiang Province},
            country={China}}

\affiliation[inst2]{organization={Key Laboratory of Coastal Environment and Resources of Zhejiang Province, School of Engineering, Westlake University,},%Department and Organization
            addressline={18 Shilongshan Road}, 
            city={Hangzhou},
            postcode={310024}, 
            state={Zhejiang Province},
            country={China}}

\affiliation[inst3]{organization={School of Engineering, Westlake University},%Department and Organization
            addressline={18 Shilongshan Road}, 
            city={Hangzhou},
            postcode={310024}, 
            state={Zhejiang Province},
            country={China}}

\begin{abstract}
Fluid-particle systems are highly sensitive to particle morphologies. While many attempts have been made on shape descriptors and coupling schemes, how to simulate the particle-particle and particle-fluid interactions with a balance between accuracy and efficiency is still a challenge, especially when complex-shaped particles are considered. This study presents a Metaball-Imaging (MI) based Discrete Element Lattice Boltzmann Method (DELBM) for fluid simulations with irregular shaped particles. The major innovation is the MI algorithm to capture the real grain shape for DELBM simulations, where the Metaball function is utilized as the mathematical representation due to its versatile and efficient expressiveness of complex shapes. The contact detection is tackled robustly by gradient calculation of the closest point with a Newton–Raphson based scheme. And the coupling with LBM is accomplished by a classic sharp-interface scheme. As for refiling, a local refiling algorithm based on the bounce back rule is implemented. Validations on three settling experiments of irregular-shaped natural cobblestones indicate the proposed model to be effective and powerful in probing micromechanics of irregular-shaped granular media immersed in fluid systems. The potential of this model on studies of shape-induced physical processes is further investigated with numerical examples on the "drafting, kissing and tumbling" phenomenon of pair particles in various shapes.

\end{abstract}

% %%Graphical abstract
% \begin{graphicalabstract}
% \includegraphics{grabs}
% \end{graphicalabstract}

% %%Research highlights
% \begin{highlights}
% \item Research highlight 1
% \item Research highlight 2
% \end{highlights}


\end{frontmatter}

%% \linenumbers

% %% The Appendices part is started with the command \appendix;
% %% appendix sections are then done as normal sections
% \appendix

%% main text
\section{Introduction}
\label{sec:intro}

Fluid-particle systems are essential in many human-related operations, including petroleum processing\cite{robinson2006petroleum}, ocean mining\cite{zhao2018experimental} and blood circulation\cite{nemmar2002passage}. In most cases, the particles involved come in a wide variety of shapes. This makes morphological features a crucial factor in the mechanical analysis of granular materials. The influence of morphologies can be appreciated at different scales producing important fluid disruptions that can be seen at the macro-scale as well as complex interactions between submerged particles at the micro-scale\cite{cho2016effects, bagheri2016drag, zhang2016lattice}. Taking an example in drug delivery, research\cite{cooley2018influence} shows that the margination and adhesion of drugs in blood vessels are highly susceptible to the drug-particle shape. From the view of experiments, the development of high-resolution imaging has provided a quantitative tool to capture the micro-structure of particles. X-ray Computed Tomography (XRCT) is one of the most prevalent techniques in this area\cite{withers2021x}. It enables 3D characterization of the studied material with high resolution. This opens up an opportunity to connect the microstructure of particles with the macro phenomena of fluid-particle systems physically\cite{xiao2019effect, shen2016characterization, marteau2021experimental}. However, how to accurately simulate the impact of morphological features is still a challenge\cite{weinhardt2021experimental}. Difficulties exist in converting the 3D XRCT images into a quantitative descriptor of morphology. In other words, complex shape features need to be compressed into a concise geometric representation. Moreover, the paradigm to probe mechanical behavior is also an indispensable component, which translates the shape descriptor into a quantitative analysis tool for particle kinematics. 

To model particle motions with complex shapes, there are basically two camps: the assembly paradigm and the avatar paradigm\cite{kawamoto2018all}. The main difference between them is how to characterize the morphological features. Once the shape is properly represented, collision detection and force calculation follow similar strategies. The assembly paradigm fits shape characteristics by clustering simple, regular graphics like spheres. For example, the sphere-clustering technique is a widely used assembly paradigm due to its simplicity\cite{garcia2009clustered, li2015multi}. It approximates the real shape by combining multiple overlapping spheres. Polyhedrons\cite{hohner2012numerical}, Ellipsoids\cite{yan2010three, regueiro2014micromorphic} and Sphere-polyhedrons\cite{dobrohotoff2012optimal, galindo2013coupled} are also used under a similar clumping framework. However, a large amount of prime graphics (For instance hundreds or thousands of triangular mesh elements as reported in the literature \cite{li2015multi,galindo2013coupled}) are needed to achieve sufficient accuracy for characterization of complex shapes especially smooth ones, which leads to a significant increase in contact detection times. This could make simulations time-costly and limit the number of involved particles. To alleviate the efficiency problem, attempts have been made on improvement of the prime graphic\cite{lim2014granular} and the contact algorithm\cite{wang2021coupled}.
But those advances still suffer from artificial surface roughness\cite{podlozhnyuk2017efficient}. 

The avatar paradigm overcomes the above dilemma by utilizing a uniform mathematical representation as the shape describer, which brings a great advantage in morphological characterization and contact detection. For instance,  Kawamoto et al\cite{kawamoto2016level} apply the level-set function in the handling of shape capture and contact detection. The proposed level-set approach achieves simulation on thousands of realistic particles reconstructed from XRCT images. The main drawbacks of it are the function interpolation and the lookup mechanism for contact detection\cite{zhao2019poly}, which leads to a heavy reliance on computer memory and performance. The superquadric-based method is another interesting implementation\cite{peng2019contact, wang2021flow,kildashti2020accurate}. It uses an extension of spheres and ellipsoids as the shape descriptor, which enables particles smooth-surfaces in simulation. But such shape expression has poor applicability for near-spherical particles\cite{zhao2020universality}. This impasse is broken by upgrading it from superquadric to poly-superellipsoid functions \cite{zhao2019poly, zhao2021sudodem}. A poly-superellipsoid is an assembly of eight separate superellipsoids, which makes it capable of characterizing more types of convex particles. The implementation of it is also more straightforward than level sets, creating  create advantages in dealing with particle motions and contact detection. However, 
% its particle contact is solved in the form of an optimization problem\cite{zhao2019poly, zhao2020universality}, which might be a burden for large-scale problems. Besides, 
poly-superellipsoid function still has some constraints on the expressed shape for now, such as symmetry and complexity\cite{peng2019contact, zhao2020universality}. This could limit their use in more complicated issues, like simulations consisting of concave sands. 

Although remarkable progress has been made in morphological characterization, it is still a challenging task to integrate realistic shapes with micro-mechanical models for probing fluid-particle interactions. The aforementioned studies mainly focus on particle-particle and particle-wall interactions. To the best of the author's knowledge, such sophisticated geometric representations have not been used in DEM simulations coupled with Computational Fluid Dynamics schemes.
% for particle-flow simulations. 

Therefore, there is still a gap in modelling realistic particle morphology with a balance between simulation accuracy and computational efficiency for fluid-particle systems. In this paper, we propose an efficient avatar paradigm called Metaball-Imaging Discrete Element Lattice Boltzmann Method (MI-DELBM), which has the ability to fully consider the morphological factors, especially the smooth and round features, in fluid-particle simulations. For characterization of realistic particles, the Metaball function\cite{uralsky2006practical} is modified and adopted as the uniform geometric descriptor. In visualization, Metaballs are n-dimensional isosurfaces. This property enables it to represent a variety of smooth, continuous surfaces, bringing advantages in shape characterization. The main innovation in this paradigm is the Metaball-Imaging algorithm, which can characterize the morphology of particles directly from XRCT images in high-resolution. This enables convincing quantitative study on shape factors and the coupled mechanical model to capture the impact of irregular shapes accurately. For probing the micro-mechanics of involved particles and fluid, a coupled Discrete Element Lattice Boltzmann Method (DELBM) is utilized. Validations on experiments show that the proposed MI-DELBM can handle complex morphologies of realistic particles in high-fidelity while infer inter-particle contact forces as well as other related quantities accurately and efficiently. 

The contents of this paper are structured as follows: Section \ref{method} is the collection of computational techniques used in this paper, which consists of three parts. Section \ref{MF} reveals the definition and properties of the used geometric descriptor. Section \ref{MI} presents a robust algorithm to characterize complex morphologies directly from XRCT images of realistic granular materials. The coupled algorithms for Metaballs to DELBM are briefly stated in Section \ref{MM}. Section \ref{cStudies} demonstrates the validation of the proposed paradigm with simulations on three settling experiments of different irregular-shaped cobblestones. The potential of MI-DELBM in study of shape impact is explored with numerical examples on the "Drafting, Kissing and Tumbling" (DKT) problem in Section \ref{applications}. In the end, Section \ref{conc} summarizes the major contributions from the present work.

\section{Methodology}\label{method}
\subsection{The Metaball Function}\label{MF}
The Metaball function is a shape descriptor for arbitrary shaped particles\cite{zhang2021metaball}.
It can express interested morphological features by an analytic form:

% \begin{equation}
%     f(\boldsymbol{x})=\sum_{i=1}^{n}  \frac{1}{1+\left(\boldsymbol{x}-\boldsymbol{\hat{x}_{i}}\right)^{2}/\sigma^2_{i}}=1
% %    \label{metaballM}
% \end{equation}

\begin{equation}
    f(\boldsymbol{x})=\sum_{i=1}^{n} \frac{\hat{k_{i}}}{\left(\boldsymbol{x}-\boldsymbol{\hat{x}_{i}}\right)^{2}}=1
    \label{metaballM}
\end{equation}
where $\boldsymbol{x}$ represents the input of position; $\boldsymbol{\hat{x}_{i}}$ is the position of control point $i$, governing the skeleton of the represented model; 
$n$ is the number of control points in this Metaball function; 
$\hat{k_i}$ is the positive coefficient determines the influence scale of $i$ th control point. By setting this function equal 1, the point $\boldsymbol{x}$ is at the contour of the represented particle surface. 

\begin{figure}[]
    \begin{centering}
    \includegraphics[width=1.0\linewidth]{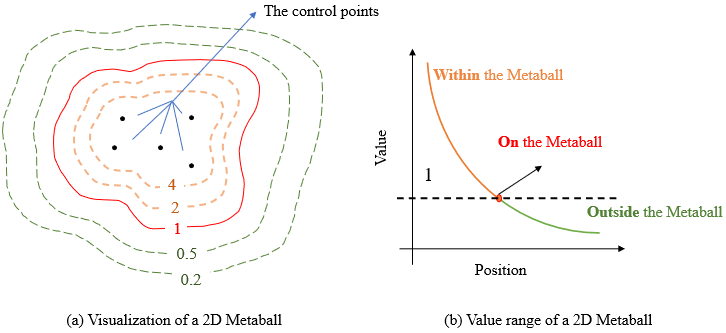}
    \caption{The visualization and value range of a 2D Metaball.}
    \label{fig:M_VR}
    \end{centering}
\end{figure}

The Metaball is an assembly of n-dimensional isosurfaces. It has no additional constraints on control points or weights, which brings it a strong applicability. Both simple and complex shapes, even with round features or concave voids, can all be properly represented. It is worth noting that the value range of the Metaball function with random input is from $0$ to $\infty$. To exemplify these advantages, a 2D Metaball is shown in Figure \ref{fig:M_VR}. The function value will be exactly 1 at the surface, with lower values signaling points outside the particle and higher values for points inside. This property is crucial in the definition of loss function for the gradient search (GS) in Metaball-Imaging, which is dedicated to generating a parametric representation of the real shape (Section \ref{sec:GS}). Furthermore, this offers simple algorithms to solve the point-inside-particle problem and the collision-detection task between DEM Metaballs. It also gives a straightforward strategy for coupling it with CFD techniques such as LBM (Section \ref{MM}), which brings advantages in efficiency. 

\subsection{The Metaball-Imaging Algorithm}\label{MI}
Here we design a Metaball-Imaging (MI) algorithm to to transform the XRCT image of irregular-shaped particle into an explicit, Metaball-function based mathematical representation, which is called {\it avatar} in this paper.
% The Metaball-Imaging (MI) Algorithm is designed to transform the XRCT image of irregular-shaped particle into an explicit, Metaball-function based mathematical representation, which is called {\it avatar} in this paper. 
The avatar possesses high-resolution morphological features and can feed the particle characterization into simulations.
% \boldsymbol{x}

This task can be simplified into an optimization problem, searching for the set of parameters $\{\hat{k_i}, \boldsymbol{\hat{x_i}}\}$ for a Metaball model which can minimize the following function:
% \begin{equation}
%     \arg \min _{k i, x i, y_{i} z_{i}}\left|\sum_{i=1}^{n} \frac{k_{i}}{\left(X-x_{i}\right)^{2}+\left(Y-y_{i}\right)^{2}+\left(Z-z_{i}\right)^{2}}-1\right|
%     \label{objF}
% \end{equation}
% where ($X,Y,Z$) = the coordinates of hull points from the CT-scan; $\{k_i,x_i,y_i,z_i\}$ = the parameter set for the learned metaball model; 
% $n$ = the number of control points. 

\begin{equation}
    \arg \min _{\hat{k_i}, \boldsymbol{\hat{x}_{i}}}\left(\sum_{i=1}^{n_1}\sum_{j=1}^{n_2} \frac{\hat{k_{i}}}{\left(\boldsymbol{x}_{j}-\boldsymbol{\hat{x}_{i}}\right)^{2}}-1\right)^2
    \label{objF}
\end{equation}
where $\{\hat{k_i}, \boldsymbol{\hat{x}_{i}}\}$ is the parameter set of the targeted Metaball model; $\boldsymbol{x}_j$ stands for the input, coordinates of hull points from XRCT; $n_1$ and $n_2$ refer to the number of control points and input samples seperately. 

Genetic Algorithm and Gradient Search are two commonly used optimization techniques for engineering problems\cite{salomon1998evolutionary}, and in this work will be used to solve Eq.~\ref{objF}. The Genetic Algorithm (GA) is a typical meta-heuristic algorithm, which refers to a series of paradigms consisting of strategies to develop a heuristic answer to an optimization problem\cite{whitley1994genetic}. It mimics the process of natural selection from a certain population. This brings it an excellent global search ability and strong robustness. GA can always achieve an optimal solution without any initial values through enough iterations, while the consuming time can be very high for complex problems. The Gradient Search (GS) is a branch of gradient-based algorithms, which can distill solutions for optimization problems by gradient of a designed objective function\cite{salomon1998evolutionary}. It originates from Gradient Descent, an efficient search strategy for local optimal solutions. GS shows excellent performance in local search with satisfying speed, while suffering from limited global search ability and strict demand for initial value.

\begin{figure}[]
    \begin{centering}
    \includegraphics[width=0.9\linewidth]{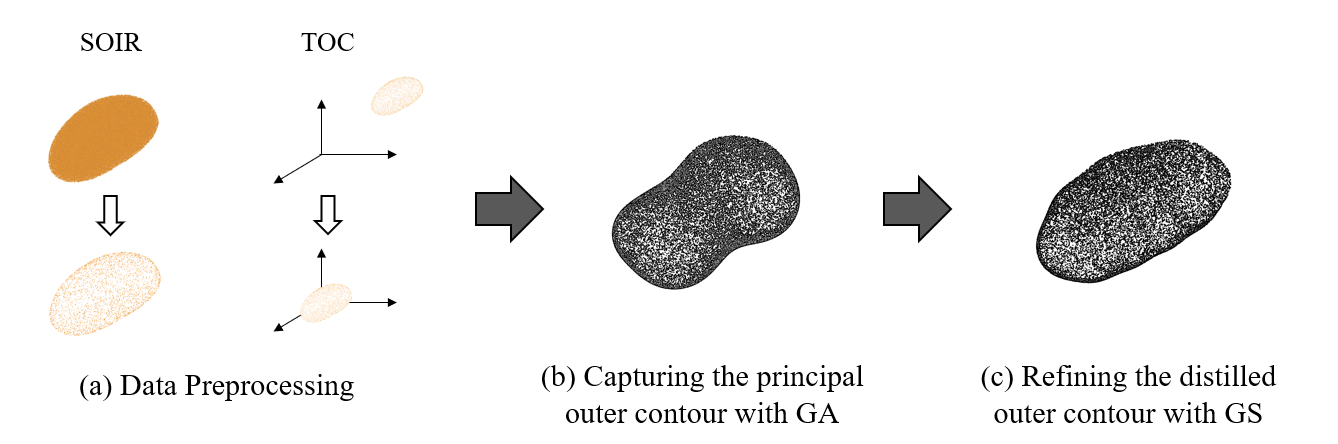}
    \caption{The algorithm flow of the Genetic-based Gradient Search algorithm.}
    \label{fig:ggs_flow}
    \end{centering}
\end{figure}

The proposed Metaball-Imaging algorithm is a combination of GA and GS, which makes the best use of their advantages and bypasses those disadvantages. In this scheme, GA is utilized to capture the principal outer contour of the targeted XRCT cloud points, searching for a sufficiently good global solution. And the obtained solution will serve as the initial value for GS. Then, GS can refine the obtained outer contour, filtering out the optimal solution by the gradient relationship. Due to the properties of meta-heuristic and gradient-based algorithms, such a framework can achieve advantages in not only accuracy but also efficiency.
% \subsubsection{The algorithm flow}\label{}
The proposed algorithm involve three steps as shown in Figure \ref{fig:ggs_flow}.

Step 1: Data preprocessing of the XRCT cloud points (Figure \ref{fig:ggs_flow} a). This preprocessing consists of two operations, the specification of interested regions (SOIR) and the transformation of coordinates (TOC). SOIR is to specify those points on outer contour from the XRCT result. This is because the fitting of the Metaball only requires the point hull. TOC is implemented to translate the specified region of interest into the coordinate system centered at the origin. Such an operation can avoid abnormal fitted parameters caused by XRCT coordinates. 
% Not only increase the fitting efficiency, but also create convenience for future
% applications. 
The point hull obtained through SOIR and TOC is noted as $\boldsymbol{H}=\{Hx_i,Hy_i,Hz_i\}$ ($i\in[1, m]$, where m represents the number of hull points).

Step 2: Capturing the principal outer contour with GA (Figure \ref{fig:ggs_flow} b). Due to GA's defect in local search and efficiency, it is hard for GA to distill a fine outer contour within an acceptable time. Considering the superiority of GA in global search, the ability to locate the range of the optimal accurately, it is used to capture the rough outer contour, i.e. a sufficiently good solution for Eq. \ref{objF}. The obtained contour then serves as the initial value in Step 3 for further refinement. It is worth noting that the point hulls in (b) and (c) of Figure \ref{fig:ggs_flow} are only for visualization. The obtained models are actually presented by Metaballs in this algorithm.

Step 3: Refining the distilled outer contour with GS (Figure \ref{fig:ggs_flow} c). GS has advantages in local search and efficiency. But its result can be easily affected by the initial value. Thus, GS is implemented to do further refinement on the principle contour distilled by GA, searching for the optimal solution for Eq. \ref{objF}. In this step, iterative calculations will be carried out based on the gradient relationship. Finally, a refined contour model will be obtained.

Step 2 and 3 will be discussed at length in Section \ref{sec:GA} and \ref{sec:GS}. All definitions are introduced in 3D forms, while they also work for 2D cases through expansion.

\subsubsection{Genetic Algorithm for capture of the principal outer contour}\label{sec:GA}
In the capture of principal outer contour by GA, five segments are included: \emph{population initialization}, \emph{mutation}, \emph{crossover}, \emph{evaluation} and \emph{selection}. 
Generations of these segments will be carried out as shown in Algorithm \ref{alg:GA}.  

\begin{algorithm}
    \caption{The Genetic Algorithm for capture of the principal outer contour}\label{alg:GA}
    
    \hspace*{\algorithmicindent} \textbf{Input:} the preprocessed point hull $\boldsymbol{H}$, the number of generations $E^{ga}$, the number of individuals in the population $N_I$, the number of genes in each individual $N_G$, the mutation coefficient $C_m$, the crossover coefficient $C_c$.\\
    \hspace*{\algorithmicindent} \textbf{Output:} the Metaball model of the principal outer contour $\boldsymbol{M}^{ga}$.
    
    \begin{algorithmic}[1]
    \State \textbf{Initialization -} $N_I$ individuals are randomly initialized with a string 
    of $N_G$ genes. The control points in each individual are set to be inside $\boldsymbol{H}$;

    \For{$i = 1,2,...,$ to $E^{ga}$}
        \State \textbf{Mutation -} For each indivudial, performing muatation with coefficient $C_m$;
        \State \textbf{Crossover -} For random pair of individuals, performing crossover with coefficient $C_c$;
        \State \textbf{Evaluation -} For each individual, calculating the fittness score;
        \State \textbf{Selection -} Selecting the fittest $N_I$ individuals for next generations;
    \EndFor

    \State \textbf{Return:} The fittest individual in the population $\boldsymbol{M}^{ga}$.
    \end{algorithmic}
\end{algorithm}

\begin{figure}[]
    \begin{centering}
    \includegraphics[width=0.9\linewidth]{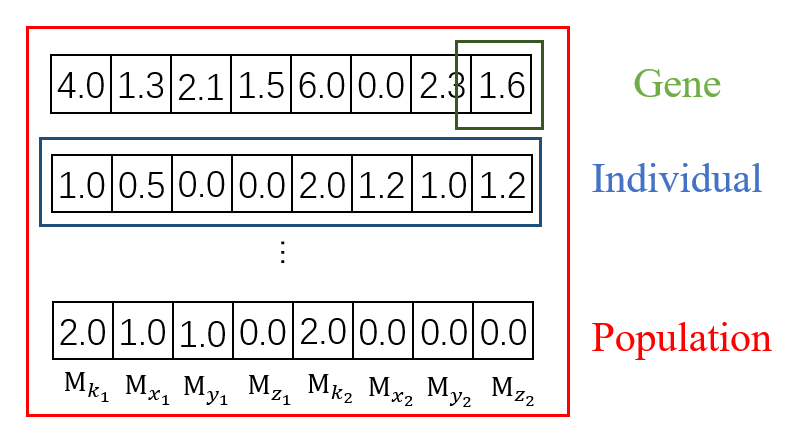}
    \caption{Population, individual and gene.}
    \label{fig:initialization}
    \end{centering}
\end{figure}

\textbf{Population Initialization}: This process refers to the initialization of N individuals as a population. Each individual represents a possible parameter set $\boldsymbol{M} = \{M_{k_i}, M_{x_i}, M_{y_i}, M_{z_i}\}$ ($i\in[1, n]$) to the fitted Metaball model (See Figure \ref{fig:initialization}). An individual consists of a series of strings defined as Genes, standing for a certain parameter in the set. The number of genes in each initialization is set to be a constant.

In this segment, all individuals are randomized with control points inside targeted point hull $H$. Such an operation is to satisfy the geometric constraint of the Metaball. The judgment of whether a point $P$ is inside $H$ is completed by linear programming. This problem is defined as the following:
\begin{equation}
\begin{array}{cl}
    \underset{A}{\operatorname{Minimize}} & \boldsymbol{C A} \\
    \text { Subject to } & \boldsymbol{H}^T \boldsymbol{A}=\boldsymbol{N}^T
\end{array}
\label{PinH}
\end{equation}

Where $\boldsymbol{C}={\underbrace{[1,1, \cdots 1]}_{m}}$; $\boldsymbol{A}={\underbrace{[a_1,a_2, \cdots a_m ]}_{m}}^T$; $\boldsymbol{H}^T=\left|\begin{array}{cccc}Hx_1, & Hx_2, & \ldots & Hx_m \\ Hy_1, & Hy_2, & \ldots & Hy_m \\ Hz_1, & Hz_2, & \cdots & Hz_m \\ 1, & 1, & \ldots & 1\end{array}\right|$, 
$\{Hx_m, Hy_m, Hz_m\}$ is the coordinate of points from the preprocessed hull $\boldsymbol{H}$;
% $m$ is the number of hull points
$\boldsymbol{N}=\left[P_{x}, P_{y}, P_{z}, 1\right]^T$, ($P_{x}, P_{y}, P_{z}$) is the coordinate of the studied point P. 

If the point $P$ is inside the studied point hull $\boldsymbol{H}$, there will 
be a solution A to Eq. \ref{PinH}, satisfying:
\begin{equation}
a_1+a_2+...+a_m = 1, a_i>0
\end{equation}

\textbf{Mutation and Crossover}: Mutation and Crossover are strategies to produce offsprings. They are the most vital segments in GA and the key to finding the optimal solution. 

\begin{figure}[]
    \begin{centering}
    \includegraphics[width=0.9\linewidth]{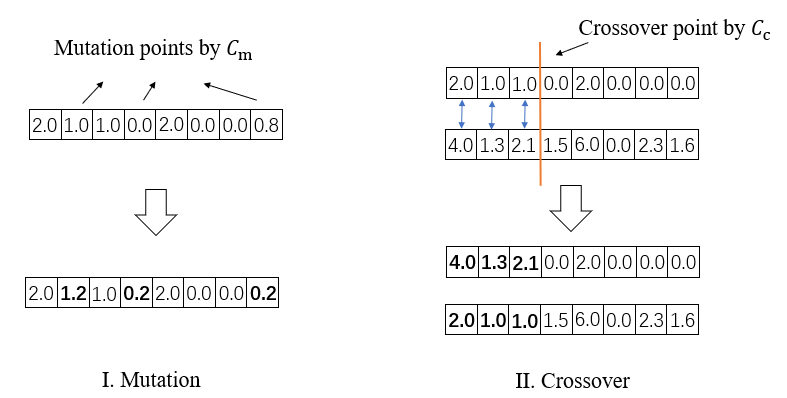}
    \caption{Mutation and Crossover.}
    \label{fig:mAc}
    \end{centering}
\end{figure}

Mutation refers to random change in the value of genes with a probability $C_m$, which stands for the change probability (Figure \ref{fig:mAc} I). This strategy is designed to control the exploration breadth and convergence rate. Crossover means the exchange of genes between two different individuals with a coefficient $C_c$, which determines the crossover point (Figure \ref{fig:mAc} II). This strategy is dedicated to filtering out better genes and promote positive diversity. A proper setting of $C_m$ and $C_c$ can maintain population diversity and prevent overfitting problems.

\textbf{Evaluation}: Evaluation involves the concept of fitness functions. It gives a fitness score, which determines the ability of this individual to compete with others in the population, to each individual. Here, the fitness function is defined as the following:
\begin{equation}
    F(\boldsymbol{M}) =\sum_{i=1}^{m}(f_{H_i}^l(M) - 1)^2
\end{equation}
where $f_{H_i}^l(M)$ = $\sum_{j=1}^{n} \frac{M_{k_j}}{\left(H_{x_i}-M_{x_j}\right)^{2}+\left(H_{y_i}-M_{y_j}\right)^{2}+\left(H_{z_i}-M_{z_j}\right)^{2}}$, $H_i$ stands for a control point in the point hull $\boldsymbol{H}$.
% Where $\{M_{k_i}, M_{x_i}, M_{y_i}, M_{z_i}\}$ = the genes of a certain individual. 

\textbf{Selection}: Selection is based on the fitness score calculated in Evaluation. It is implemented to filter the fittest N individuals and pass them to the next generation. It is generally performed after mutation, crossover and evaluation. This makes individuals with higher fitness more likely to survive and reproduce.

\subsubsection{Gradient Search for refinement of outer contour}\label{sec:GS}
In the refinement of outer contour by GS, two segments are included in GS: \emph{Gradient Descent} and \emph{the anomaly detection}. The flowchart of GS is detailed in Algorithm \ref{alg:GS}.  

\begin{algorithm}
    \caption{The Gradient Search for refinement of outer contour}\label{alg:GS}
    
    \hspace*{\algorithmicindent} \textbf{Input:} the preprocessed point hull $H$, the number of generations $E^{gs}$, the learning rate $\eta$, the Metaball model of the principal outer contour $\boldsymbol{M}^{ga}$.\\   
    \hspace*{\algorithmicindent} \textbf{Output:} the metaball model for the refined outer contour $\boldsymbol{M}^{gs}$.
    
    \begin{algorithmic}[1]
    \State $\boldsymbol{M}^{ga}$ is taken as the inital parameter $\boldsymbol{M}_0$; 
    % \State $\theta \leftarrow P_0$;
    \For{$i = 1,2,...,$ to $E^{gs}$} \Comment{ \textbf{1st Gradient Descent} }
        \State $\boldsymbol{M}_{0} \leftarrow \boldsymbol{M}_{0}  -\eta \cdot \nabla_{\boldsymbol{M}_{0}} L(\boldsymbol{M}_{0})$;
    \EndFor

    \State \textbf{Anomaly Detection - } Performing clearning on $M_0$ for two types of the anomaly points: control point overflow and sign abnormality. 
    Sending the cleaned parameters $M_1$ into the next step;
    
    \For{$i = 1,2,...,$ to $E^{gs}$} \Comment{ \textbf{2nd Gradient Descent} }
    \State $\boldsymbol{M}_{1} \leftarrow \boldsymbol{M}_{1}  -\eta \cdot \nabla_{\boldsymbol{M}_{1}} L(\boldsymbol{M}_{1})$;
    \EndFor

    \State \textbf{Return:} The searched paramter set $\hat{\theta}$.
    \end{algorithmic}

\end{algorithm}

\textbf{Gradient Descent}:
Gradient Descent is an optimization algorithm based on gradient information, which is readily available from the Metaball functions. For an objective function $L(M)$, its parameters can be updated iteratively to find the optimal:
\begin{equation}
    \boldsymbol{M} \leftarrow \boldsymbol{M} -\eta \cdot \nabla_{\boldsymbol{M} } L(\boldsymbol{M} )
\end{equation}
where $\boldsymbol{M}$ represents the parameters for Gradient Descent, here is the parameter set of the fitted Metaball model; $\eta$ is the learning rate; $\nabla_{\boldsymbol{M}} L(\boldsymbol{M} )$ is the gradient of the $L(\boldsymbol{M} )$ to the parameter $\boldsymbol{M}$. 

\begin{figure}[]
    \begin{centering}
    \includegraphics[width=0.5\linewidth]{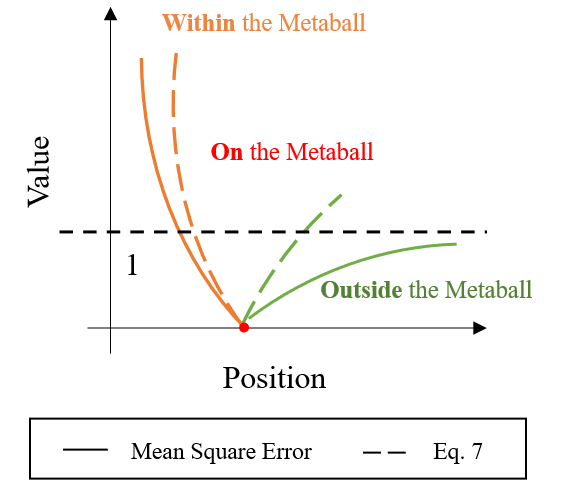}
    \caption{The value range of the objective function with Mean Square Error and Eq. \ref{GSloss} form. The latest one increases the loss function value for points outside the point hull, improving the GS performance.}
    \label{fig:mseLoss}
    \end{centering}
\end{figure}

In this segment, the objective function for contour refinement is defined as a piecewise function instead of Eq. \ref{objF}:

\begin{equation}
    L(\boldsymbol{M} )= \begin{cases}\sum_{i=1}^{m}(f_{H_i}^l(\boldsymbol{M})-1)^{2}, & f_{H_i}^l \in[2,+\infty) \\ \sum_{i=1}^{m}|f_{H_i}^l(\boldsymbol{M})-1|, & f_{H_i}^l \in[1,2] \\ 
    \sum_{i=1}^{m}\left[(f_{H_i}^l(\boldsymbol{M})-1)^{2}+\frac{1}{f_{H_i}^l(\boldsymbol{M})}-1\right], & f_{H_i}^l \in[0,1]\end{cases}
    \label{GSloss}
\end{equation}
% where $\alpha$ is chosen to be 10 in this study, although experimentation has showed that this value should be increased for very complex shapes. 
This is because many attempts have shown that a loss function in Mean Square Error form can often result in distorted models with control points outside the targeted hull. This is related to the property of Metaball function. When the study point is internally close to or externally far from the Metaball hull, its corresponding function value will all be very small. This results in the value range of Eq. \ref{objF} as shown by lines in Figure \ref{fig:mseLoss}, which makes the direct GS fall into the local optimal solution easily. This defect can be avoided through the implementation of Eq. \ref{GSloss}. Under this form, the loss value of points outside and close to the Metaball hull can all be enlarged greatly (Dash lines in Figure \ref{fig:mseLoss}). Such implementation can not only improve fitting efficiency but also adaptability to complex geometry of GS.

% This definition is from the property of Metaball function. If the loss function is defined here simply by Mean Square Error form, the value range of it will be as shown by lines in Figure \ref{fig:mseLoss}. When the study point is internally close to or externally far from the Metaball hull, its corresponding function value will all be very small. This will result in negative impacts on GS due to its reliance on gradient relations. Distorted models with control points outside the targeted hull is very likely to be obtained. To avoid this, Eq. \ref{GSloss} is adopted. This equation can greatly enlarge the loss value of points outside or close to the Metaball hull (Dash lines in Figure \ref{fig:mseLoss}). Such implementation can not only improve the fitting efficiency but also adaptability to complex geometry of the GS.

For gradient update, we adopt a pattern that the gradient for the whole dataset, the entire point hull H, will be calculated once for each round of iteration. This is because sufficient attention to all points on the targeted hull can endow the trained model with higher fidelity\cite{ruder2016overview}.

\textbf{Anomaly Detection}:
Since the search of GS is strictly based on the gradient relationship, redundant control points can be merged reasonably in this process. But this also raises two other problems: control point overflow and sign abnormality. The control point overflow refers to solution with control points outside the target point hull $H$. And the sign abnormality means solution with control points of negative $k$ values. Those anomaly points will be cleared out by assigning a zero weight during this stage.

\subsection{Micro-mechanical Models for Numerical Simulations with Metaballs}\label{MM}
% It takes a similar framework as described in (Metaball-DELBM can't be searched). A sharp interface coupling scheme is utilized.
% Many special algorithms
% are also proposed to reduce numerical noise.
% Details are omitted here due to length reasons. 

To probe the mechanical behavior of the avatar distilled by MI, DEM and LBM are implemented with Metaballs. The coupling takes a similar scheme as stated in \cite{zhang2021metaball, https://doi.org/10.48550/arxiv.2206.11634}, thus only key points are included here for the sake of brevity.

\subsubsection{Coupling with DEM model}
DEM is a method tackling motions and effects of large number of individual particles\cite{cundall1992numerical, luding2008introduction}. In this framework, the translation is expressed by Newton's equation while the rotation is described the angular momentum conservation model:
\begin{equation}
\left\{\begin{array}{l}
m_i \boldsymbol{a}_i=m_i \boldsymbol{g}+\sum_{j=1}^N \boldsymbol{F}_{i j}^c+\boldsymbol{F}_i^e \\
\frac{d}{d t}\left(\mathbf{I}_i \boldsymbol{\omega}_i\right)=\sum_{j=1}^N \boldsymbol{T}_{i j}^c+\boldsymbol{T}_i^e
\end{array}\right.
\end{equation}
where $m_i$ and $\boldsymbol{a_i}$ stand for the mass and acceleration of the i-th particle. $\boldsymbol{F}_i^e$ is the external force. $\boldsymbol{F}_{i j}$ means the contact force between the i-th particle and its neighbor particle j. $I_i$ is the inertia tensor. $\omega_i$ is the angular velocity. ${T}_{i j}^c$ and $\boldsymbol{T}_i^e$ represent torques caused by the external and contact force seperately. More details can be referenced in \cite{zhang2021metaball}

Here, the contact force $F^c$ is described by the linear spring dashpot model\cite{cundall1992numerical}. The normal component of $F^c$ is given by:$\mathrm{F}_n^c=k_n \delta+\eta_n\left(\boldsymbol{v}_{\boldsymbol{j}}-\boldsymbol{v}_{\boldsymbol{i}}\right) \cdot \boldsymbol{n}$, where $k_n$ is the normal spring stiffness. $\delta$ is the particle overlapping distance. $\eta_n$ is the normal damping coefficient. And $\boldsymbol{n}$ stands for the unit normal vector. 

The tangential component of $F^c$ is determined as:
\begin{equation}
\begin{cases}F_t^c & =\min \left(\mu_s F_n^c,F_{t 0}^c\right)\\F_{t 0}^c & =\left\|-k_t \boldsymbol{\xi}-\eta_t\left(\boldsymbol{v}_{\boldsymbol{j}}-\boldsymbol{v}_{\boldsymbol{i}}\right) \cdot \boldsymbol{t}\right\|\end{cases}
\end{equation}
where $\mu_s$ stands for the static friction coefficient. $k_t$ is the tangential spring stiffness. $\eta_t$ refers to the damping coefficient. $\boldsymbol{\xi}$ is the tangential spring. And $\boldsymbol{t}$ is the unit tangential vector. 

For DEM coupling, the collisions between Metaballs as well as Metaball and wall are handled properly. This requires three parameters: the contact point $x_{cp}$, the contact direction $\bm{n}$ and the overlap $\delta$. To avoid unwanted intersection problems,the collision is first transformed to a sphero-Metaball based contact\cite{zhang2021metaball}, where the original Metaball is approximated by a combination of zoomed internal Metaball and dilated sphere with radius $R_s$. Then, an optimization problem is introduced to locate the closest points on studied Metaball set or Metaball and wall. This optimization problem is defined as:
\begin{equation}
\begin{array}{cl}
    {\operatorname{Minimize}} & f_0(\bm{x})+f_1(\bm{x}) \\
    \text { Subject to } &  c_{tol}<\abs{f_0(\bm{x})}<1, \quad c_{tol}<f_1(\bm{x})<1
\end{array}
\label{CM_coreF}
\end{equation}

where the $f_0(\bm{x})$, $f_1(\bm{x})$ = functions of two Metaballs; $c_{tol}$ = the tolerance to avoid $f_0(\bm{x})+f_1(\bm{x})=0$ when $\norm{\bm{x}}\to \infty$. Under this limitation, 
the distilled solution will make the gradient of Eq.~\ref{CM_coreF} equal to 0:
\begin{equation}
\nabla (f_0(\bm{x})+f_1(\bm{x}))=\bm{0}
\label{eq:grad}
\end{equation} 

On collisions between Metaballs, a Newton-Raphson method is used to search the local minimum point $x_m$. And the closest points on Metaballs ($x_{c0}$ and $x_{c1}$) can be approximated by:
\begin{equation}
\begin{cases}
& \bm{x}_{c0} = \bm{x}_m + q_0\nabla f_0(\bm{x}_m) \\
& \bm{x}_{c1} = \bm{x}_m + q_1\nabla f_1(\bm{x}_m)
\end{cases}
\label{eq:cp}
\end{equation}

By combining Taylor series expansion and $f_{0}(x_{c0})$ = 1, $q_0$ and $q_1$ can be expressed explicitly to 
calculate the $x_{c0}$ and $x_{c1}$. Then, the required three parameters for collision between Metaballs can be defined as:
\begin{equation}
\begin{cases}
& \bm{x}_{cp} = \bm{x}_{c0} + (R_{s0}-0.5\delta)\bm{n}\\
& \bm{n} = \frac{\bm{x}_{c0} - \bm{x}_{c1}}{\norm{\bm{x}_{c0} - \bm{x}_{c1}}} \\
& \delta = R_{s0}+R_{s1}-\norm{\bm{x}_{c1} - \bm{x}_{c0}}
\end{cases}
\label{eq:contactprop}
\end{equation}

On collisions between Metaball and wall, the problem is simplified into finding a point $x^{W}_{cw}$ on the wall with normalized gradient of $(-1, 0, 0)$ to rotated Metaball function. This point is searched similarly by a Newton-Raphson method. Then the closest point on Metaball to wall is defined as:
\begin{equation}
\begin{aligned}
\bm{x}^{W}_{cp} & = \bm{x}^{W}_{cw} + q^W\nabla f^{W}(\bm{x}^{W}_{cw})
\end{aligned}
\label{eq:metawallcp}
\end{equation}

With $f^{W}(\bm{x}^{W}_{cw})$ = 1 and Taylor series expansion, the required three parameters for collision between Metaball 
and wall can be defined as:
\begin{equation}
\begin{cases}
& \bm{x}_{cp} = \bm{x}_{cw} + 0.5\delta\bm{n}\\
& \bm{n} = \frac{\bm{x}_{cm} - \bm{x}_{cw}}{\norm{\bm{x}_{cm} - \bm{x}_{cw}}} \\
& \delta = R_{s}-\norm{\bm{x}_{cm} - \bm{x}_{cw}} 
\end{cases}
\label{eq:contactprop2}
\end{equation}

% More theoretical and implementation details can be found in \cite{zhang2021metaball}.

\subsubsection{Coupling with LBM model}
LBM is a popular Computational Fluid Dynamics scheme based on kinetic theory. Instead of solving flows at the continuum scale, the Boltzmann equation is used to describe fluid motions, in which the distribution functions are the basic variable. The distribution function $G_{i}(\bm{x},t)$ representing the probability that a molecule of fluid occupies a cell at position $\bm{x}$ with a given discrete velocity $\bm{e_i}$ at time $t$. More details on how Navier-Stokes equations can be solved by LBM can be found in \cite{galindo2013coupled}. 

\begin{figure}[]
    \begin{centering}
    \includegraphics[width=0.6\linewidth]{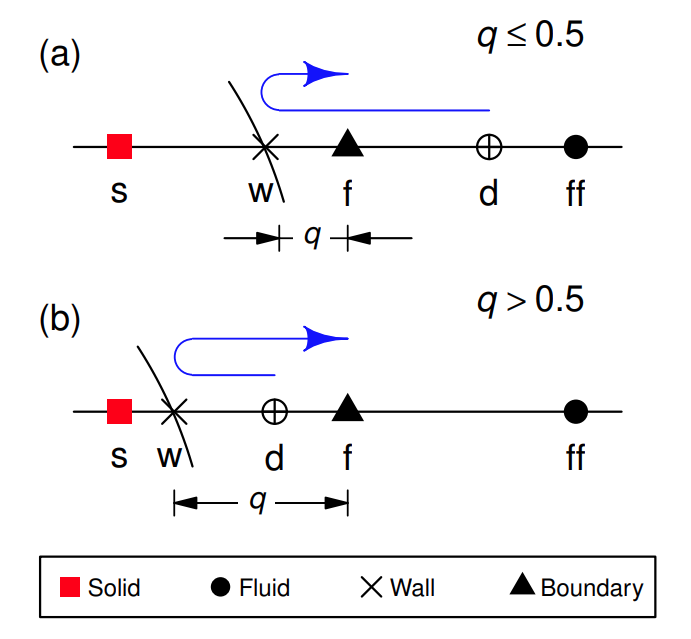}
    \caption{The schematic of interpolated bounce-back rule.}
    \label{fig:IBB}
    \end{centering}
\end{figure}

For DEM-LBM coupling, the boundary condition, hydrodynamic force and refilling algorithm are selected carefully. On moving boundary condition, the interpolated bounce-back (IBB)\cite{peng2008comparative} is deployed to capture the influence of Metaball particles on the fluid and impose the non-slipping boundary condition. This scheme divides the involved nodes into fluid nodes, solid nodes and boundary nodes (fluid nodes that are next to the solid boundary)\cite{https://doi.org/10.48550/arxiv.2206.11634}. These nodes are symbolized with "s" - the closest solid node, "w" - the wall, "f" - the boundary node, and "ff" for the neighbouring fluid node of "f" as shown in Figure \ref{fig:IBB}. Under this definition, the distribution function of a point which departs from $x_d$ with velocity $e_{i'}$ ($i'$ is the direction opposite to $i$), hits on the wall and returns back to $x_f$ with velocity $e_i$ is defined as:
\begin{equation}
G_{i}(\bm{x}_f,t+\Delta{t}_{LBM}) = G^{+}_{i'}(\bm{x}_d,t) + 6\omega_{i'}\rho_{f}\frac{\bm{e}_i \cdot \bm{u}_{w}}{C^2},
\label{eq:eqneq1}
\end{equation}
where $\omega_i$ are the LBM weights for each discrete direction; $\rho_{f}$ represents the fluid density;
the particle surface velocity $\bm{u}_{w} = \bm{v}_{pj} + \bm{w}_{pj} \times (\bm{x}_w-\bm{x}_{pj})$, where $\bm{v}_{pj}$ and $\bm{w}_{pj}$ are the translational and angular velocity at the $j$th particle's centroid $\bm{x}_{pj}$, respectively. $G^{+}_{i'}(\bm{x}_d,t)$ is the post-collision distribution function and can be decomposed into equilibrium and non-equilibrium parts:
\begin{equation}
G^{+}_{i'}(\bm{x}_d,t) = G^{eq}_{i'}(\rho_d,\bm{u}_d) + G^{neq}_{i'}(\bm{x}_d,t),
\label{eq:eqneq2}
\end{equation}

The $\bm{u}_d$ in above equation can be expressed explicitly by linear interpolation:
\begin{equation}
\bm{u}_d = \frac{1}{3}\bm{u}^{*}_d + \frac{2}{3}\bm{u}^{**}_d.
\label{eq:neqd}
\end{equation}
where $\bm{u}^{*}_d = \left\{ 
  \begin{array}{l l}
    \text{$2q\bm{u}_f + (1-2q)\bm{u}_{ff}$}, & \quad \text{$q \leqslant 0.5$,}\\[3ex]
     \text{$\frac{1-q}{q}\bm{u}_f + \frac{2q-1}{q}\bm{u}_w$}, & \quad \text{$q > 0.5$,}\\[0.ex]
\end{array} \right.$, $q$ can be obtained by solving $f(\bm{x}_f + q\bm{e}_i)$ = $c_0$, and $c_0$ is a Metaball function value depending on $R_s$; $\bm{u}^{**}_d = \frac{1-q}{1+q}\bm{u}_{ff} + \frac{2q}{1+q}\bm{u}_w$. More details on this IBB scheme can be found in \cite{https://doi.org/10.48550/arxiv.2206.11634}.

On hydrodynamic forces, a momentum exchange method (MEM)\cite{chen2013momentum} is utilized.
MEM is a frequently used scheme for fluid-particle interactions, where the hydrodynamic force is defined as the sum of the momentum exchanges among all discrete velocity colliding with solid surfaces. To obey the Galilean invariance principle,  a Galilean-invariance MEM\cite{lorenz2009corrected}, which can efficiently avoid numerical noises, is implemented. 

As a DEM particles vacates LBM cells, some missing $G_i$ comming out of the particle must be initialized. To achieve this, a local refilling algorithm based on bounce-back rule is applied. The reinitialized distribution function is defined as: 
\begin{equation}
G_{i}(\bm{x}_{new}, t) = G_{i'}(\bm{x}_{new}, t) + 6\omega_{i'}\rho_{f}\frac{\bm{e}_i \cdot \bm{u}_{w}}{C^2},
\end{equation}
where $\bm{x}_{new}$ is the new fluid node position. If $G_{i'}(\bm{x}_{new}, t)$ does not exist, the equilibrium refilling is used: $G_{i}(\bm{x}_{new}, t) = G^{eq}(\rho_0, \bm{u}_{new})$, where $\bm{u}_{new} = \bm{v}_{pj} + \bm{w}_{pj} \times (\bm{x}_{new}-\bm{x}_{pj})$. 

% Details of the above coupling scheme can be referenced in \cite{https://doi.org/10.48550/arxiv.2206.11634}. 

\section{Validation on Particle Settling}\label{cStudies}
To validate the proposed algorithm, we conducted simulations on the settling of avatars distilled from XRCT images taken on three different irregular-shaped natural cobblestones in various viscous fluids. Comparisons are made on important indicators for this physical process.

\begin{figure}
    % \begin{centering}
    \includegraphics[width=1.0\linewidth]{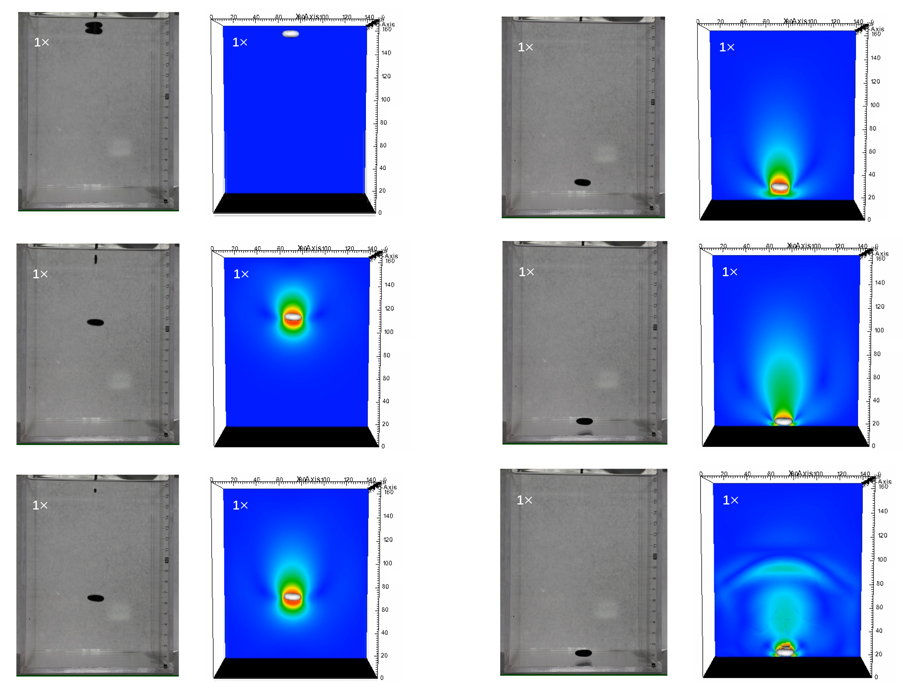}
    \caption{The experiment and simulation snapshots of irregular-shaped cobblestone-settling in viscous fluid.}
    \label{fig:ES}
    % \end{centering}
\end{figure}

The experimental setup is shown in Fig \ref{fig:ES} as well as simulations snapshots for one to one comparison. The experiments are carried
out in a rectangular container with outside dimension: 0.15$\times$0.15$\times$0.20$m$ and
internal dimension: 0.14$\times$0.14$\times$0.195$m$.

\begin{figure}[]
    \begin{centering}
    \includegraphics[width=1.0\linewidth]{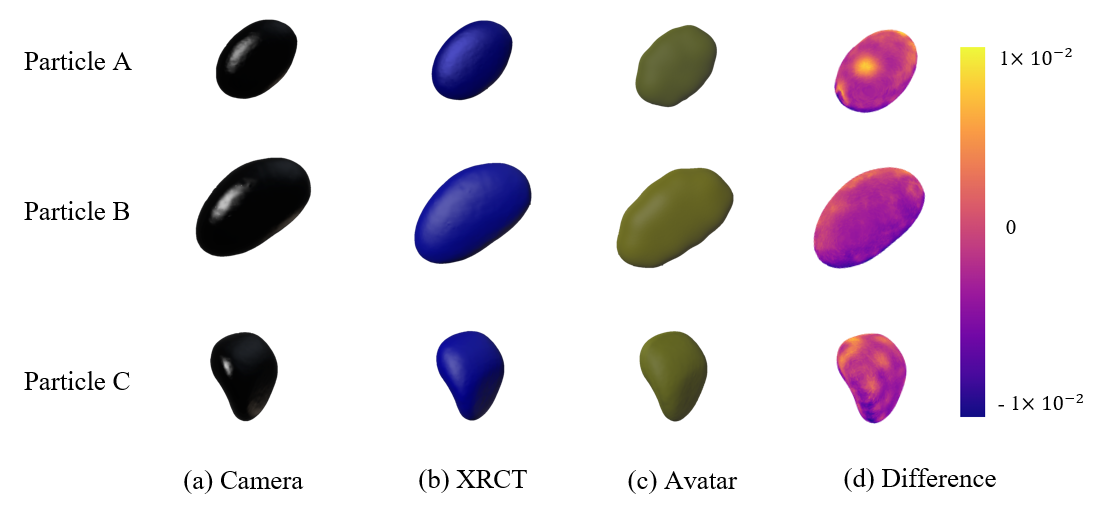}
    \caption{Visualization of the experimental cobblestones (a) High-speed camera images (b) XRCT images (c) The Metaball avatar (d) Difference between XRCT and avatar - the color stands for the loss value of each hull point from XRCT on the Metaball avatar in Eq. \ref{GSloss}, the average loss values of these particles are 6.1$\times 10^{-3}$ - particle A, 4.1$\times 10^{-3}$ - particle B and 3.8$\times 10^{-3}$ - particle C.}
    \label{fig:RD} 
    \end{centering}
\end{figure}

The shapes of these cobblestones are shown in Figure \ref{fig:RD} (a), which are irregular and different. The densities and volumes are 2850.67$kg/m^3$ and 6.314$\times 10^{-7} m^3$ for cobble $A$,  2559.838$kg/m^3$ and 9.063$\times 10^{-7} m^3$for
cobble $B$ as well as 2957.951$kg/m^3$ and 6.119$\times 10^{-7} m^3$ for cobble $C$. Three types of dimethyl silicone oil
with viscosity 0.088, 0.295 and 0.853 $Pa \cdot s$ are used as the viscous fluid in experiments(Measured by LICHEN NDJ viscometer). The cobblestones are released by tweezers in a completely submerged state. The initial release orientation is set to make the maximum projection area as perpendicular to the settling direction as possible. The setting trajectory of cobblestones are recorded by a high-speed camera (CANON EOS 5D MARK IV). The settling velocity is extracted with MATLAB by the pixel change of cobblestone centroid between video frames.

The XRCT image, for shape digitization, is captured with a ZEISS Xradia 610 Versa. The voxel size is 45.476311$\mu m$ and optical magnification 0.395810. The particle in the simulation contains $7.974538 \times 10^6$ voxels in average to represent the real geometry. The particle segmentation is done by "ilastik"\cite{sommer2011ilastik}, an edge detection algorithm for segmentation of XRCT images aided by machine learning. Different from function methods, "ilastik" detects particle edges by considering both voxel intensity and brightness gradient. 
This avoids omitting particles with concave spots on the surface or broken spots on the edge\cite{lei2018pore}. 
The segmented particles are shown in Figure \ref{fig:RD}, (b).

To make one to one comparisons, the settings of simulations are identical to the experiments. Furthermore, no parameters are
calibrated in the simulation since the MI-DELBM is a fully solved scheme, the only free parameter is the fluid viscosity which is measured independently. The MI avatars are produced with the following hyperparameter settings: $E^{ga} = 100$, $N_I = 1200$, $N_G = 100$, $C_m = 0.6$, $C_c = 0.5$, $E^{gs} = 100000$ and $\eta = 0.001$, the shapes of these cobblestones are captured by the MI algorithm as avatars shown in Figure \ref{fig:RD} (c). Comparisons (Figure \ref{fig:RD} d) indicate MI to be a reliable tool to capture particle morphology. It is worth noting that the surface mesh is only for visualization and those particles are presented as Metaballs in the simulation system. The LBM and DEM time steps are set as: 2.0$\times10^{-4}$ and 2.0$\times10^{-6}s$. The LBM space step is set as 1$\times 10^{-3}m$.
In the following discussion, the Reynold number is defined uniformly as: 
\begin{equation}
    Re = \frac{\rho_f U_p D_e}{\mu}
\end{equation}
Where $\rho_f$ = the density of fluid; $U_p$ = the peak velocity of the particle; $D_e$ = the diameter 
of the equal volume sphere; $\mu$ = the viscosity of the fluid. 

\begin{figure}[]
    \begin{centering}
    \includegraphics[width=0.8\linewidth]{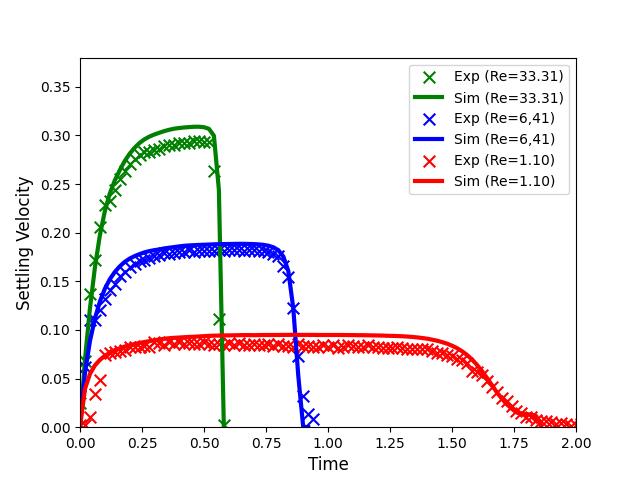}
    \caption{The comparison of vertical velocity between the simulation and experiment for Particle A during the settling.}
    \label{fig:v_m88}
    \end{centering}
\end{figure}

\begin{figure}[]
    \begin{centering}
    \includegraphics[width=0.8\linewidth]{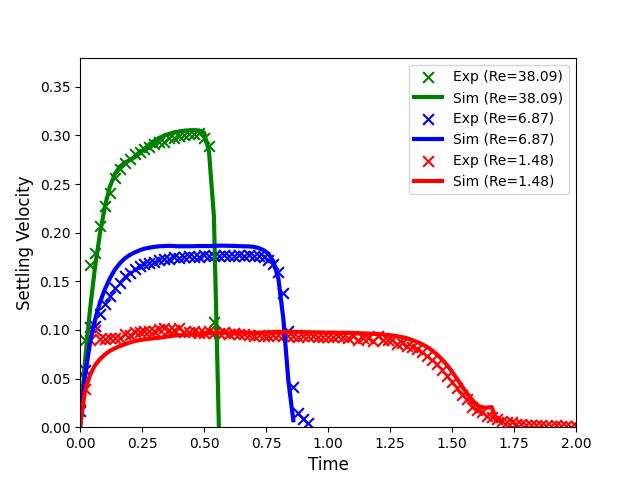}
    \caption{The comparison of vertical velocity between the simulation and experiment for Particle B during the settling.}
    \label{fig:v_m132}
    \end{centering}
\end{figure}

\begin{figure}[]
    \begin{centering}
    \includegraphics[width=0.8\linewidth]{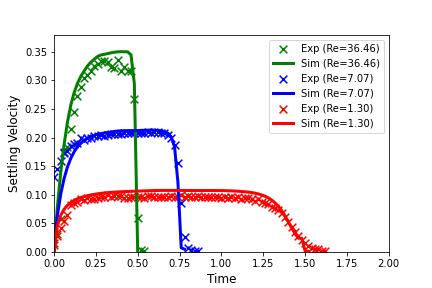}
    \caption{The comparison of vertical velocity between the simulation and experiment for Particle C during the settling.}
    \label{fig:v_m133}
    \end{centering}
\end{figure}

Overall, the simulation results match well with the experiments for all $Re$  considered. Figure \ref{fig:v_m88}, \ref{fig:v_m132} and \ref{fig:v_m133} show the vertical-velocity time series for particle A, B and C in both experiments and simulations respectively. For particle A, little rotations are observed in the experiments which were not observed in the simulations. However, the settling velocities match. For particle B, small deviations can be observed in the initial stage at Re = 1.48 and 6.87. This is related to the fact that the initial releasing orientation of particle B cannot be controlled strictly due to its geometric complexity. Therefore, particle B needs to adjust itself at the initial stage to maximize its projection area and this causes these deviations. For particle C, many rotations are observed in experiments, especially at Re = 36.46. Regarding geometry, particle C has the most complex shape, which makes its initial release orientation hard to satisfy for the perpendicular condition. This explains the small deviation in the initial stage for particle C at Re = 1.30 and 7.07. As for Re = 36.46, the relatively high Re makes the particle vulnerable to disturbance. This leads to a more intense rotation for orientation adjustment in the ultimate stage, which causes the curve fluctuation. However, the error of the steady settling velocity is all within an acceptable range, which further proves the robustness of the proposed method. 

% 结果描述

\section{Application of the proposed methodology on the "Drafting, Kisssing, and Tumbling" (DKT) Phenomenon of Pair Particles}\label{applications}
Particle shapes can have significant influences on particle-settling behaviors\cite{zhang2016lattice}. MI-DELBM provides us with an efficient method to study the impact of morphological parameters on physical processes.
To demonstrate its capability, we conduct a series of simulations on settling of pair particles with various shapes in viscous liquid. It is well known that settling of pair spheres follows the "drafting, kissing and tumbling" (DKT) process\cite{wang2014drafting}: the following particle will surpass the leading particle due to the drag reduction from the wake of the leading particle. However, how particle shape influences the DKT process is not fully explored. Therefore, we set all these particles with the same density 2800 $kg/m^{3}$ and volume 1e-6 $m^{3}$.
Simulations are carried out in a rectangular container with dimension: 0.20$\times$0.20$\times$0.30$m$ ($W$=0.20$m$, $H$=0.30$m$).
The fluid density and viscosity are 976 $kg/m^{3}$ and 0.104 $m^{2}/s$. The pair particles are placed at 0.270 and 0.255 $m$ from bottom initially, with a horizontal distance of 0.004 $m$. 
To study the impact of shapes quantitatively, four morphological parameters are introduced, namely, the sphericity ($\Phi$), nominal
diameter ($d_n$), surface-equivalent-sphere diameter ($d_s$), and Corey Shape Factor ($CSF$).

The sphericity ($\Phi$) is a frequently used shape factor for non-spherical particles and is defined as:
\begin{equation}
    \Phi=\frac{S_{\text {sphere }}}{S}
\end{equation}
where $S_{\text {sphere }}$ = the surface area of the volume-equivalent sphere to the studied particle; $S$ = the surface area 
of the studied particle. 

Another two widely used parameters are nominal diameter $d_n$ and surface-equivalent-sphere diameter $d_s$. The $d_n$ is defined as the diameter of the volume-equivalent sphere. And the $d_s$, 
\begin{equation}
    d_{s}=\sqrt{\frac{4 A_{p}}{\pi}}
\end{equation}
where $A_{p}$ = the projected area of the particle. 

The $CSF$ reveals the dimension feature of the studied particle, as given by
\begin{equation}
    \mathrm{CSF}=\frac{d_{s}}{\sqrt{d_{i} d_{l}}}
\end{equation}

\begin{figure}[]
    \begin{centering}
    \includegraphics[width=0.8\linewidth]{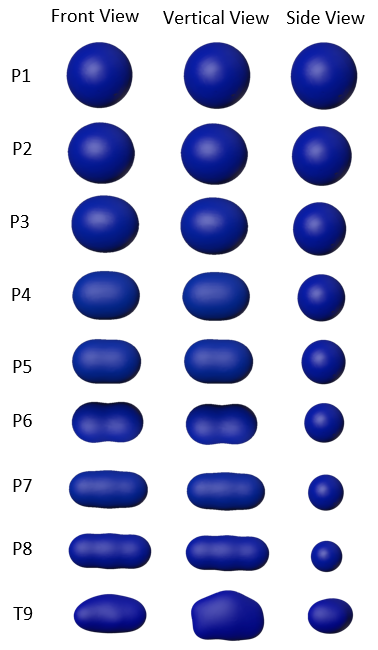}
    \caption{The seven set of particles with stepped morphological parameters.}
    \label{fig:particles}
    \end{centering}
\end{figure}

Since limited particles can be obtained from XRCT due to cost problems, the involved particles, $P_1$ to $P_8$, are designed manually to have stepped morphological parameters, of which $CSF$ is the main varying parameter (Table \ref{T:MP_particles}). A lower $CSF$ means higher degree of difference. In visualization, control points of these particles are moved away gradually (Figure \ref{fig:particles}).It is worth noting that $P1$ is the volume equivalent sphere to the rest of the involved particles. The diameter of $P1$ is $D_e$ = 0.0125 $m$. We also select a cobblestone particle, $T_9$, which has the same shape features of $P_8$ except the sphericity. The MI avatar of it is shown in Figure \ref{fig:particles}. It can be observed that the difference in sphercity brings higher degree of anisotropy into $T_9$, compared to $P_8$.

\begin{table}[]
    \centering
    \caption{The morphological parameters of the particles in DKT}
    \begin{tabular}{|c|c|c|c|}
    \hline
    Particle & Sphericity & dn/ds  & CSF    \\ \hline
    P1       & 1.0000     & 1.0000 & 1.0000 \\ \hline
    P2       & 1.0000     & 0.9845 & 0.9563 \\ \hline
    P3       & 0.9937     & 0.9718 & 0.9173 \\ \hline
    P4       & 0.9812     & 0.9514 & 0.8535 \\ \hline
    P5       & 0.9633     & 0.9335 & 0.8002 \\ \hline
    P6       & 0.9440     & 0.9172 & 0.7521 \\ \hline
    P7       & 0.9225     & 0.8959 & 0.7075 \\ \hline
    P8       & 0.8966     & 0.8780 & 0.6648 \\ \hline
    T9       & 0.9460     & 0.8738 & 0.6663 \\ \hline
    \end{tabular}
    \centering
    \label{T:MP_particles}
\end{table}

The results show that non-spherical particle motions follow similar patterns as spheres. It also suggests that morphological parameters play an important role in DKT process. Henceforth, the vertical position is normalized by $H$, time by $t_r$ = $\sqrt{H/g}$,
velocity by $\sqrt{Hg}$, angular velocity by $2\pi/t_r$ and distance between particles by $D_e$.

% Henceforth, the vertical position is normalized by container height 0.3 $m$, time by $\sqrt{0.3 m / 9.8 m/s^2} =  0.175s$,
% velocity by $\sqrt{0.3 m \times 9.8 m/s^2} = 1.715m/s$, angular velocity by $2\pi/0.175s = 35.90$ 

\begin{figure}
    \begin{centering}
    \includegraphics[width=0.8\linewidth]{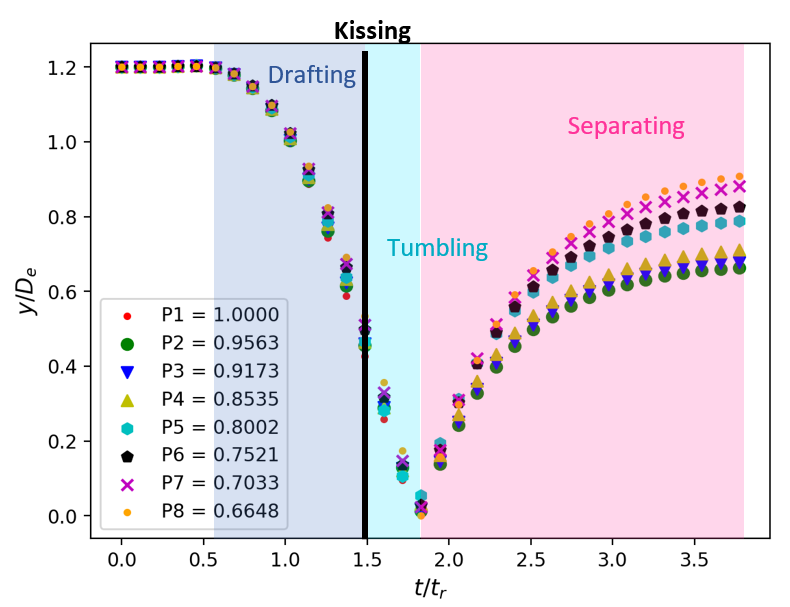}
    \caption{The vertical distance between particles of different morphological parameters, the number in legend stands for CSF value of each particle.  }
    \label{fig:vd}
    \end{centering}
\end{figure}

\begin{figure}
    \begin{centering}
    \includegraphics[width=1.0\linewidth]{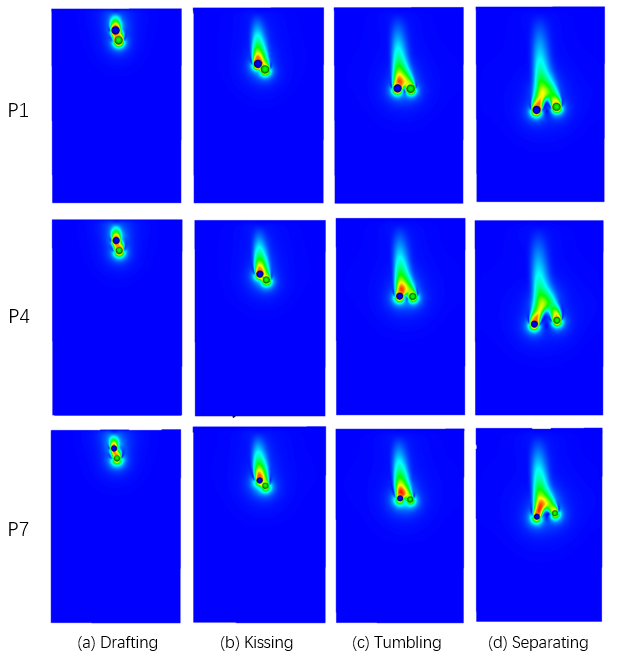}
    \caption{The snapshot of the DKT simulations for P1, P4 and P7 from the side view at the same time-step. Color indicates the fluid velocity magnitude.}
    \label{fig:snapshot}
    \end{centering}
\end{figure}

Figure \ref{fig:vd} indicates the change of the vertical distance between the leading and the following particles.
It can be observed that the $CSF$ is in a negative correlation with the vertical distance between particles during the DKT process. This phenomenon is related to the flow field induced by the morphology and interaction of particles (Figure \ref{fig:snapshot}). From flow patterns of P1, P4 and P7, the increase in CSF can obviously enhance the flow field of the following particles and weaken that of the leading particles. Such an impact is more significant in the separation stage. While this correlation is not obvious before drafting and after separating since the particle interaction is not yet formed at that time. 

\begin{figure}[]
    \begin{centering}
    \includegraphics[width=1.\linewidth]{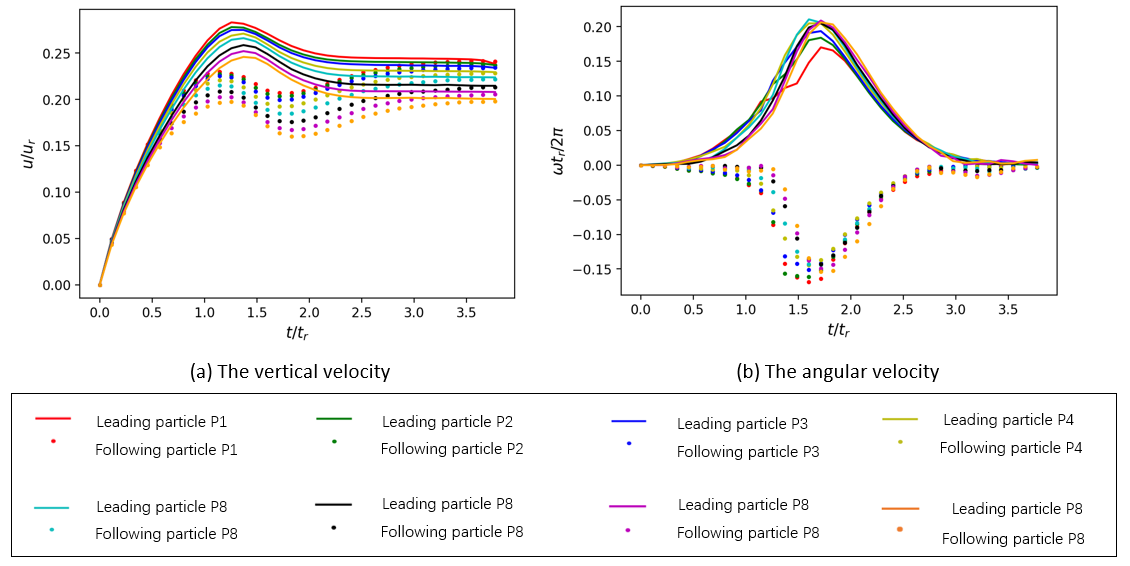}
    \caption{The time series of vertical velocity and angular velocity of particle pairs $P_1$ to $P_8$.}
    \label{fig:vv_av}
    \end{centering}
\end{figure}

\begin{figure}
    \begin{centering}
    \includegraphics[width=0.8\linewidth]{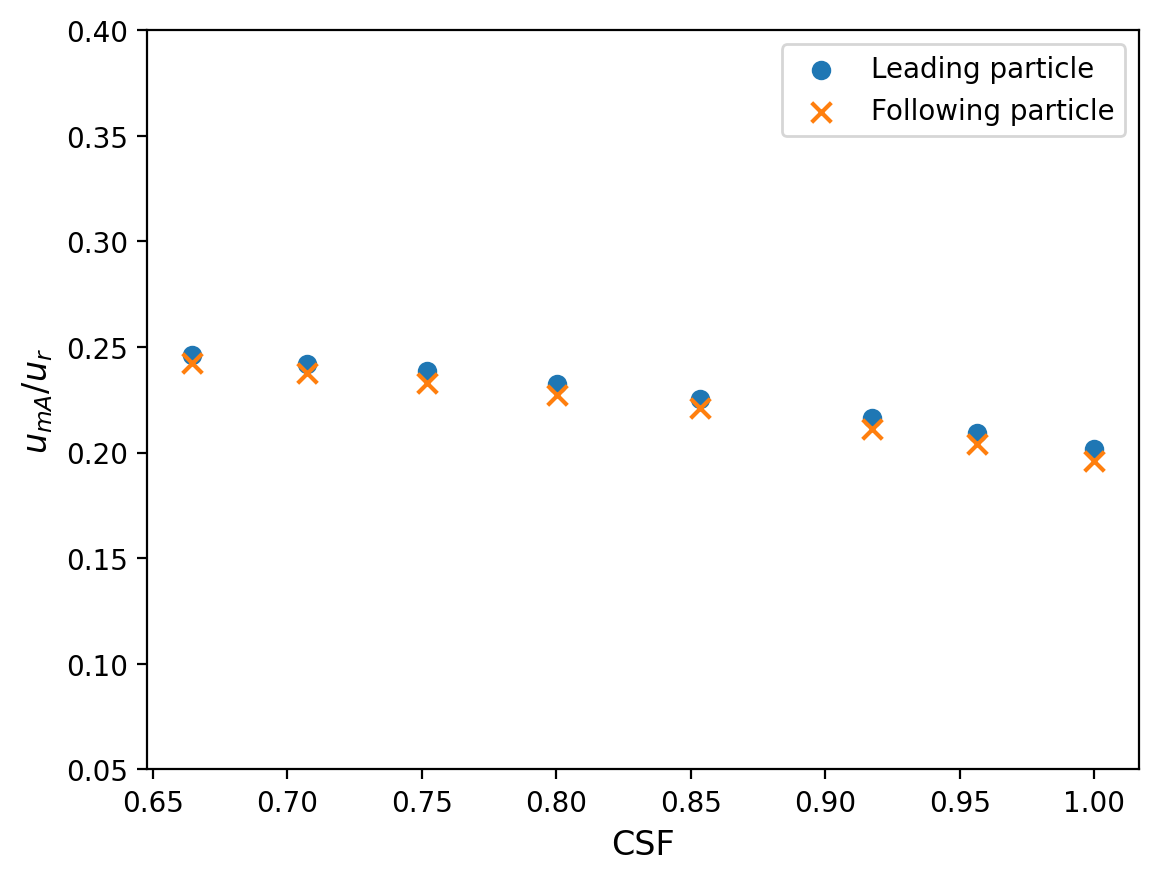}
    \caption{The settling velocity of particle pairs $P_1$ to $P_8$.}
    \label{fig:V}
    \end{centering}
\end{figure}

\begin{figure}
    \begin{centering}
    \includegraphics[width=0.8\linewidth]{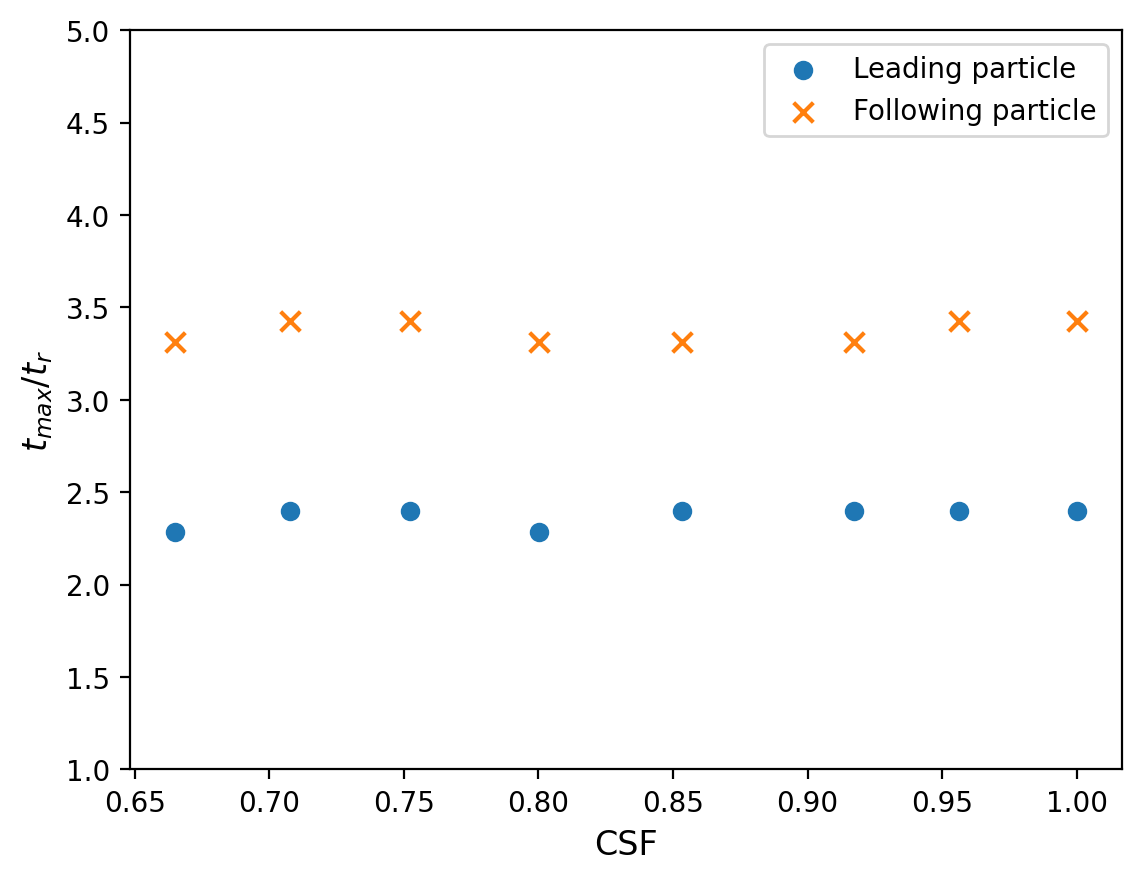}
    \caption{The time of peak velocity of particle pairs $P_1$ to $P_8$.}
    \label{fig:V_t }
    \end{centering}
\end{figure}

\begin{figure}
    \begin{centering}
    \includegraphics[width=0.8\linewidth]{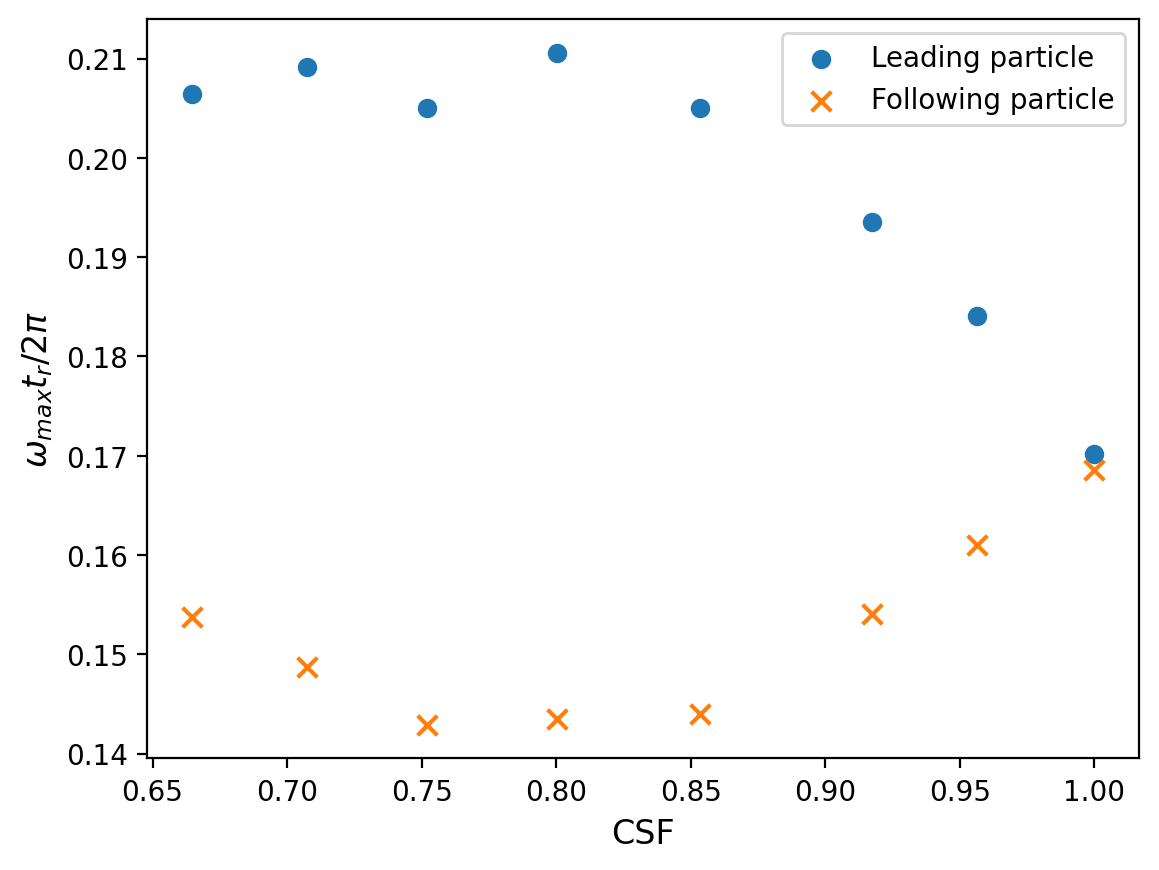}
    \caption{The maximum angular velocity of particle pairs $P_1$ to $P_8$.}
    \label{fig:MAV}
    \end{centering}
\end{figure}

\begin{figure}
    \begin{centering}
    \includegraphics[width=0.8\linewidth]{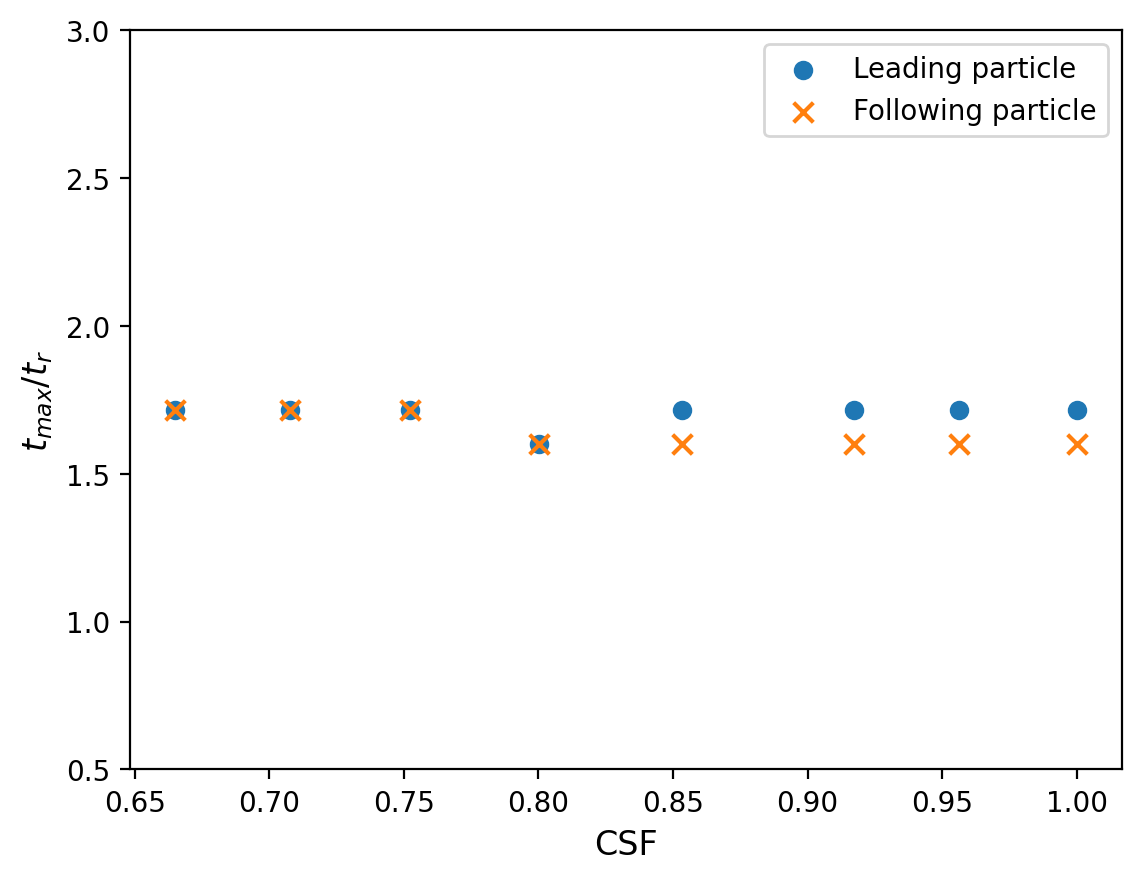}
    \caption{The time of maximum angular velocity of particle pairs $P_1$ to $P_8$.}
    \label{fig:MAV_time}
    \end{centering}
\end{figure}

Figure \ref{fig:vv_av} reveals the time series for the vertical velocity and angular velocity magnitude of these studied particles. For settling velocity, it is clear that the decrease in $CSF$ has a negative impact on the settling velocity of both the leading and following particles. As $CSF$ ranges from $0.66$ to $1.00$, the settling velocity of both particles gradually decreases(Figure \ref{fig:V}). This is related to the impact of $CSF$ on the particle flow field. The enhancement in the wake of the following particle leads to a negative impact on the settling velocity of the leading particle. It is also known that the particle with lower $CSF$ tends to have larger maximum projected area under the same volume\cite{zhang2016lattice, liu2021interaction}. And the maximum projected area is in negative correlation with the settling velocity. These factors playing together cause the above phenomenon. However, the change in $CSF$ has little impact on the time of peak velocity for pair-particle settling (Figure \ref{fig:V_t }, $t_{mU}$ = the time of peak velocity). For angular velocity, the impact of CSF is obvious on the magnitude (Figure \ref{fig:MAV}). The rotation in DKT problem is mainly caused by particle interaction, which continues throughout the settling. With $CSF$ ranges from 0.66 to 1.00, the magnitude of the leading particles decreases gradually (Figure \ref{fig:MAV}). As for the following particles, the magnitude first decreases then increases. This indicates that the increase in anisotropy degree can lead to a stronger rotation. Surprisingly, the sum of the angular velocity magnitudes for both leading and following particles is about the same, around 0.35 $\pm$ 0.03 $1/s$ in our simulations. This reveals that the overall angular momentum of the pair particles remains basically unchanged. Similar to results of settling velocity, $CSF$ has little impact on the time required to reach the maximum velocity (Figure \ref{fig:MAV_time}, $t_mA$ = the time of maximum angular velocity). Particles with different $CSF$ reach peak angular magnitude basically at the same time. This means that the difference in shapes won't interfere with the rotation sequence of pair-particle settling. In conclusion, particle morphology has a strong impact on the settling dynamics of pair particles. Such impact is not limited in translation but also obvious in rotation. While the settling process is not sensitive to the shape change. 
% This could be caused by the interaction between particles. 

\begin{figure}
    \begin{centering}
    \includegraphics[width=0.8\linewidth]{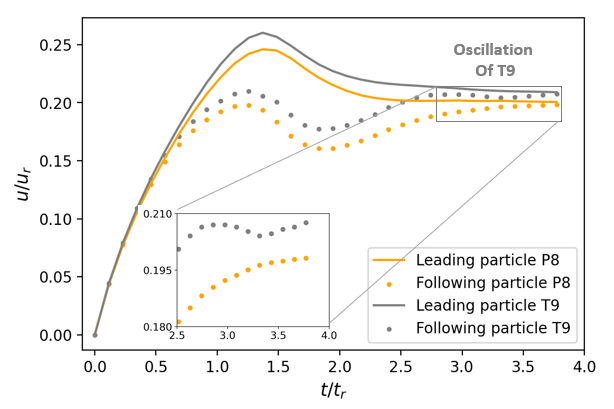}
    \caption{The settling velocity of particle pairs P8 and T9}
    \label{fig:v_p8t9}
    \end{centering}
\end{figure}

\begin{figure}
    \begin{centering}
    \includegraphics[width=0.8\linewidth]{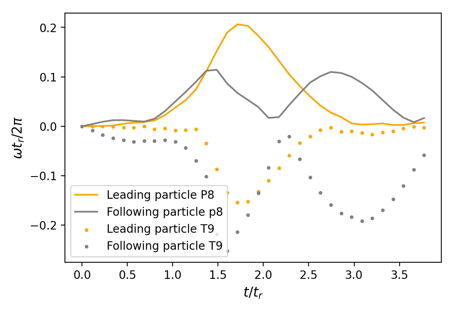}
    \caption{The angular velocity of particle pairs P8 and T9}
    \label{fig:a_p8t9}
    \end{centering}
\end{figure}

On this basis, we further compare the DKT process of particle pairs $P_8$ and $T_9$, which have the same shape features except the sphericity. For settling velocity, obvious difference can be observed on these two particles(Figure \ref{fig:v_p8t9}). Although they share the same $CSF$ and $d_n/d_s$, $T_9$ particle pair shows a larger settling velocity and different settling pattern. This phenomenon is mainly caused by the difference in maximum projected area caused by sphericity. The higher sphericity brings more flatness into $T_9$, which results in larger projected area and higher degree of anisotropy. Note that $T_9$ particle pair go through an oscillation process during separation, which cannot be observed on the rest particles. This is induced by the interaction from the leading particle to the following one, which makes the following particle deviate from the steady settling state. Together with the flat feature, more rotations are needed to adjust its posture for settling with the maximum projected area. This is in consistency with the results of angular velocity. Figure \ref{fig:a_p8t9} illustrates the angular velocities of particle pairs $P_8$ and $T_9$. Unlike the others, the $T_9$ particle pair possesses two velocity peaks. The velocity magnitude of the following particle is also stronger, especially during the separation, which indicates an oscillation process. The above finding further reveals that shape features play an important role in the DKT process and single shape feature is insufficient to characterize such impact. For more practical study, a combination of shape features are needed. 

\section{Conclusions}\label{conc}
The particle shape has been a crucial factor for the fluid-particle system. To study it quantitatively, we propose the MI-DELBM, which can simulate the fluid-particle system directly with input of
XRCT images. Implementations of Metaball-Imaging for avatar capturing and coupled micromechanical models for physical simulations are explained in detail. 

To validate the proposed model, simulations of avatars with morphological factors captured from XRCT images of different irregular-shaped particles are carried out on a series of particle settling experiments without calibration. The comparisons are in a good match. This shows that the MI-DELBM can simulate the particle-particle and fluid-particle interactions of complex shapes accurately.

Finally, we explore the potential of MI-DELBM on the classic DKT process of pair-particle settling. A quantitative study is carried out with CSF as the major variable. It is found that the shape does have a strong impact on the sinking and rotation of particles in the DKT process. This impact focuses on the magnitude of settling and angular velocity. 
With CSF ranging from 0.66 to 1.00, the peak velocity of leading and following particles increases gradually. As for angular velocity, the magnitude of leading particle first increases then decreases while the following particle shows an opposite behavior. The morphology has little influence on dynamic sequence. All 8 sets of particles reach magnitude of settling and angular velocity at almost the same time.
It is worth noting that single shape feature cannot accurately reflect the impact of shape on the DKT process. A feature combination is needed.

In conclusion, the above results reveal that the proposed MI-DELBM model is an efficient tool to fill the gap between experiments and simulations on fluid-particle systems with complex, irregular-shaped particles. Potential extensions of it can be found in not only settling behavior in industrial processing but also transportation processes such as pollutant diffusion, dense suspension and drug absorption.

\section*{Acknowledgement}
We gratefully acknowledge the funds from National Natural Science Foundation of China (Project No.12172305), Natural Science Foundation of Zhejiang Province, China (LHZ21E090002), Key Research and Development Program of Zhejiang Province (Grant No.2021C02048) and Westlake University. The presented simulations were conducted based on the ComFluSoM open source library (https://github.com/peizhang-cn/ComFluSoM).

\section*{Data Availability}
The data that support the findings of this study are available from the corresponding author upon reasonable request.

%% If you have bibdatabase file and want bibtex to generate the
%% bibitems, please use
%%
\bibliographystyle{elsarticle-num} 
\bibliography{mybib}

\begin{thebibliography}{10}
\expandafter\ifx\csname url\endcsname\relax
  \def\url#1{\texttt{#1}}\fi
\expandafter\ifx\csname urlprefix\endcsname\relax\def\urlprefix{URL }\fi
\expandafter\ifx\csname href\endcsname\relax
  \def\href#1#2{#2} \def\path#1{#1}\fi

\bibitem{robinson2006petroleum}
P.~R. Robinson, Petroleum processing overview, in: Practical advances in
  petroleum processing, Springer, 2006, pp. 1--78.

\bibitem{zhao2018experimental}
G.~Zhao, L.~Xiao, T.~Peng, M.~Zhang, Experimental research on hydraulic
  collecting spherical particles in deep sea mining, Energies 11~(8) (2018)
  1938.

\bibitem{nemmar2002passage}
A.~Nemmar, P.~M. Hoet, B.~Vanquickenborne, D.~Dinsdale, M.~Thomeer,
  M.~Hoylaerts, H.~Vanbilloen, L.~Mortelmans, B.~Nemery, Passage of inhaled
  particles into the blood circulation in humans, Circulation 105~(4) (2002)
  411--414.

\bibitem{cho2016effects}
J.~Cho, H.~Y. Sohn, Effects of particle shape and size distribution on the
  overall fluid-solid reaction rates of particle assemblages, The Canadian
  Journal of Chemical Engineering 94~(8) (2016) 1516--1523.

\bibitem{bagheri2016drag}
G.~Bagheri, C.~Bonadonna, On the drag of freely falling non-spherical
  particles, Powder Technology 301 (2016) 526--544.

\bibitem{zhang2016lattice}
P.~Zhang, S.~Galindo-Torres, H.~Tang, G.~Jin, A.~Scheuermann, L.~Li, Lattice
  boltzmann simulations of settling behaviors of irregularly shaped particles,
  Physical Review E 93~(6) (2016) 062612.

\bibitem{cooley2018influence}
M.~Cooley, A.~Sarode, M.~Hoore, D.~A. Fedosov, S.~Mitragotri, A.~S. Gupta,
  Influence of particle size and shape on their margination and wall-adhesion:
  implications in drug delivery vehicle design across nano-to-micro scale,
  Nanoscale 10~(32) (2018) 15350--15364.

\bibitem{withers2021x}
P.~J. Withers, C.~Bouman, S.~Carmignato, V.~Cnudde, D.~Grimaldi, C.~K. Hagen,
  E.~Maire, M.~Manley, A.~Du~Plessis, S.~R. Stock, X-ray computed tomography,
  Nature Reviews Methods Primers 1~(1) (2021) 1--21.

\bibitem{xiao2019effect}
Y.~Xiao, L.~Long, T.~Matthew~Evans, H.~Zhou, H.~Liu, A.~W. Stuedlein, Effect of
  particle shape on stress-dilatancy responses of medium-dense sands, Journal
  of Geotechnical and Geoenvironmental Engineering 145~(2) (2019) 04018105.

\bibitem{shen2016characterization}
W.~Shen, Z.~Yang, L.~Cao, L.~Cao, Y.~Liu, H.~Yang, Z.~Lu, J.~Bai,
  Characterization of manufactured sand: Particle shape, surface texture and
  behavior in concrete, Construction and Building materials 114 (2016)
  595--601.

\bibitem{marteau2021experimental}
E.~Marteau, J.~E. Andrade, An experimental study of the effect of particle
  shape on force transmission and mobilized strength of granular materials,
  Journal of Applied Mechanics 88~(11) (2021).

\bibitem{weinhardt2021experimental}
F.~Weinhardt, H.~Class, S.~Vahid~Dastjerdi, N.~Karadimitriou, D.~Lee, H.~Steeb,
  Experimental methods and imaging for enzymatically induced calcite
  precipitation in a microfluidic cell, Water Resources Research 57~(3) (2021)
  e2020WR029361.

\bibitem{kawamoto2018all}
R.~Kawamoto, E.~And{\`o}, G.~Viggiani, J.~E. Andrade, All you need is shape:
  predicting shear banding in sand with ls-dem, Journal of the Mechanics and
  Physics of Solids 111 (2018) 375--392.

\bibitem{garcia2009clustered}
X.~Garcia, J.-P. Latham, J.-s. XIANG, J.~Harrison, A clustered overlapping
  sphere algorithm to represent real particles in discrete element modelling,
  Geotechnique 59~(9) (2009) 779--784.

\bibitem{li2015multi}
C.-Q. Li, W.-J. Xu, Q.-S. Meng, Multi-sphere approximation of real particles
  for dem simulation based on a modified greedy heuristic algorithm, Powder
  Technology 286 (2015) 478--487.

\bibitem{hohner2012numerical}
D.~H{\"o}hner, S.~Wirtz, V.~Scherer, A numerical study on the influence of
  particle shape on hopper discharge within the polyhedral and multi-sphere
  discrete element method, Powder technology 226 (2012) 16--28.

\bibitem{yan2010three}
B.~Yan, R.~A. Regueiro, S.~Sture, Three-dimensional ellipsoidal discrete
  element modeling of granular materials and its coupling with finite element
  facets, Engineering Computations (2010).

\bibitem{regueiro2014micromorphic}
R.~Regueiro, B.~Zhang, F.~Shahabi, K.~Soga, Micromorphic continuum stress
  measures calculated from three-dimensional ellipsoidal discrete element
  simulations on granular media, IS-Cambridge 2014 (2014) 1--6.

\bibitem{dobrohotoff2012optimal}
P.~B. Dobrohotoff, S.~I. Azeezullah, F.~Maggi, F.~Alonso-Marroquin, Optimal
  description of two-dimensional complex-shaped objects using spheropolygons,
  Granular Matter 14~(5) (2012) 651--658.

\bibitem{galindo2013coupled}
S.~Galindo-Torres, A coupled discrete element lattice boltzmann method for the
  simulation of fluid--solid interaction with particles of general shapes,
  Computer Methods in Applied Mechanics and Engineering 265 (2013) 107--119.

\bibitem{lim2014granular}
K.-W. Lim, J.~E. Andrade, Granular element method for three-dimensional
  discrete element calculations, International Journal for Numerical and
  Analytical Methods in Geomechanics 38~(2) (2014) 167--188.

\bibitem{wang2021coupled}
M.~Wang, Y.~Feng, T.~Qu, T.~Zhao, A coupled polygonal dem-lbm technique based
  on an immersed boundary method and energy-conserving contact algorithm,
  Powder Technology 381 (2021) 101--109.

\bibitem{podlozhnyuk2017efficient}
A.~Podlozhnyuk, S.~Pirker, C.~Kloss, Efficient implementation of superquadric
  particles in discrete element method within an open-source framework,
  Computational Particle Mechanics 4~(1) (2017) 101--118.

\bibitem{kawamoto2016level}
R.~Kawamoto, E.~And{\`o}, G.~Viggiani, J.~E. Andrade, Level set discrete
  element method for three-dimensional computations with triaxial case study,
  Journal of the Mechanics and Physics of Solids 91 (2016) 1--13.

\bibitem{zhao2019poly}
S.~Zhao, J.~Zhao, A poly-superellipsoid-based approach on particle morphology
  for dem modeling of granular media, International Journal for Numerical and
  Analytical Methods in Geomechanics 43~(13) (2019) 2147--2169.

\bibitem{peng2019contact}
D.~Peng, K.~J. Hanley, Contact detection between convex polyhedra and
  superquadrics in discrete element codes, Powder Technology 356 (2019) 11--20.

\bibitem{wang2021flow}
S.~Wang, S.~Ji, Flow characteristics of nonspherical granular materials
  simulated with multi-superquadric elements, Particuology 54 (2021) 25--36.

\bibitem{kildashti2020accurate}
K.~Kildashti, K.~Dong, B.~Samali, An accurate geometric contact force model for
  super-quadric particles, Computer Methods in Applied Mechanics and
  Engineering 360 (2020) 112774.

\bibitem{zhao2020universality}
S.~Zhao, J.~Zhao, N.~Guo, Universality of internal structure characteristics in
  granular media under shear, Physical Review E 101~(1) (2020) 012906.

\bibitem{zhao2021sudodem}
S.~Zhao, J.~Zhao, Sudodem: Unleashing the predictive power of the discrete
  element method on simulation for non-spherical granular particles, Computer
  Physics Communications 259 (2021) 107670.

\bibitem{uralsky2006practical}
Y.~Uralsky, Practical metaballs and implicit surfaces, in: Game Developper
  Conference, 2006.

\bibitem{zhang2021metaball}
P.~Zhang, Y.~Dong, S.~Galindo-Torres, A.~Scheuermann, L.~Li, Metaball based
  discrete element method for general shaped particles with round features,
  Computational Mechanics 67~(4) (2021) 1243--1254.

\bibitem{salomon1998evolutionary}
R.~Salomon, Evolutionary algorithms and gradient search: similarities and
  differences, IEEE Transactions on Evolutionary Computation 2~(2) (1998)
  45--55.

\bibitem{whitley1994genetic}
D.~Whitley, A genetic algorithm tutorial, Statistics and computing 4~(2) (1994)
  65--85.

\bibitem{ruder2016overview}
S.~Ruder, An overview of gradient descent optimization algorithms, arXiv
  preprint arXiv:1609.04747 (2016).

\bibitem{https://doi.org/10.48550/arxiv.2206.11634}
P.~Zhang, L.~Qiu, S.~A. Galindo-Torres, Y.~Chen, A.~Scheuermann, L.~Li,
  \href{https://arxiv.org/abs/2206.11634}{Coupled metaball discrete element
  lattice boltzmann method for fluid-particle systems with non-spherical
  particle shapes: A sharp interface coupling scheme} (2022).
\newblock \href {https://doi.org/10.48550/ARXIV.2206.11634}
  {\path{doi:10.48550/ARXIV.2206.11634}}.
\newline\urlprefix\url{https://arxiv.org/abs/2206.11634}

\bibitem{cundall1992numerical}
P.~A. Cundall, R.~D. Hart, Numerical modelling of discontinua, Engineering
  computations (1992).

\bibitem{luding2008introduction}
S.~Luding, Introduction to discrete element methods: basic of contact force
  models and how to perform the micro-macro transition to continuum theory,
  European journal of environmental and civil engineering 12~(7-8) (2008)
  785--826.

\bibitem{peng2008comparative}
Y.~Peng, L.-S. Luo, A comparative study of immersed-boundary and interpolated
  bounce-back methods in lbe, Progress in Computational Fluid Dynamics, an
  International Journal 8~(1-4) (2008) 156--167.

\bibitem{chen2013momentum}
Y.~Chen, Q.~Cai, Z.~Xia, M.~Wang, S.~Chen, Momentum-exchange method in lattice
  boltzmann simulations of particle-fluid interactions, Physical Review E
  88~(1) (2013) 013303.

\bibitem{lorenz2009corrected}
E.~Lorenz, A.~Caiazzo, A.~G. Hoekstra, Corrected momentum exchange method for
  lattice boltzmann simulations of suspension flow, Physical Review E 79~(3)
  (2009) 036705.

\bibitem{sommer2011ilastik}
C.~Sommer, C.~Straehle, U.~Koethe, F.~A. Hamprecht, Ilastik: Interactive
  learning and segmentation toolkit, in: 2011 IEEE international symposium on
  biomedical imaging: From nano to macro, IEEE, 2011, pp. 230--233.

\bibitem{lei2018pore}
L.~Lei, Y.~Seol, K.~Jarvis, Pore-scale visualization of methane hydrate-bearing
  sediments with micro-ct, Geophysical Research Letters 45~(11) (2018)
  5417--5426.

\bibitem{wang2014drafting}
L.~Wang, Z.~Guo, J.~Mi, Drafting, kissing and tumbling process of two particles
  with different sizes, Computers \& Fluids 96 (2014) 20--34.

\bibitem{liu2021interaction}
J.~Liu, P.~Zhang, Y.~Xiao, Z.~Wang, S.~Yuan, H.~Tang, Interaction between dual
  spherical particles during settling in fluid, Physics of Fluids 33~(1) (2021)
  013312.

\end{thebibliography}

%% else use the following coding to input the bibitems directly in the
%% TeX file.

% \begin{thebibliography}{00}

% %% \bibitem{label}
% %% Text of bibliographic item

% \bibitem{}

% \end{thebibliography}
\end{document}